\documentclass[fleqn,usenatbib]{mnras}

\usepackage{newtxtext}
\usepackage[varvw]{newtxmath}

\usepackage[T1]{fontenc}

\DeclareRobustCommand{\VAN}[3]{#2}
\let\VANthebibliography\thebibliography
\def\thebibliography{\DeclareRobustCommand{\VAN}[3]{##3}\VANthebibliography}

\usepackage{graphicx}	

\usepackage{soul}

\def\refjnl#1{{\rm#1}}
\def\aj{\refjnl{AJ}}                   
\def\apj{\refjnl{ApJ}}                 
\def\apjl{\refjnl{ApJ}}                
\def\apjs{\refjnl{ApJS}}               
\def\ao{\refjnl{Appl.~Opt.}}           
\def\aap{\refjnl{A\&A}}                
\def\mnras{\refjnl{MNRAS}}             
\def\pasj{\refjnl{PASJ}}               

\def\be{\begin{equation}} 
\def\ee{\end{equation}}

\def\msun{{\Msun}}

\def\HI{\hbox{H~$\scriptstyle\rm I\ $}}

\def\gsim{\lower.5ex\hbox{\gtsima}} 
\def\lsim{\lower.5ex\hbox{\ltsima}} \def\gtsima{$\; \buildrel > \over 
\sim \;$} \def\ltsima{$\; \buildrel < \over \sim \;$} \def\prosima{$\; 
\buildrel \propto \over \sim \;$} \def\gsim{\lower.5ex\hbox{\gtsima}} 
\def\lsim{\lower.5ex\hbox{\ltsima}} 
\def\simgt{\lower.5ex\hbox{\gtsima}} 
\def\simlt{\lower.5ex\hbox{\ltsima}} 
\def\simpr{\lower.5ex\hbox{\prosima}} \def\la{\lsim} \def\ga{\gsim} 
  
 \def\gtsima{$\; \buildrel > \over \sim \;$} 
\def\ltsima{$\; \buildrel < \over \sim \;$} 
\def\gsim{\lower.5ex\hbox{\gtsima}} 
\def\lsim{\lower.5ex\hbox{\ltsima}} 
\def\simgt{\lower.5ex\hbox{\gtsima}} 
\def\simlt{\lower.5ex\hbox{\ltsima}} 
\def\simpr{\lower.5ex\hbox{\prosima}} 
\def\la{\lsim} 
\def\ga{\gsim}

\def\msun{\,{\rm \Msun}}

\def\E3{{\cal E}_{\rm g}^{III}}

\def\Msun{\rm M_\odot}

\def\Msun{\rm M_\odot}

\def\M*{M_*}

\def\Z*{Z_*}
\def\L*{L_*}

\title[LAEs in the EoR]{Astraeus VIII: A new framework for Lyman-$\alpha$ emitters applied to different reionisation scenarios}

\author[Hutter et al.]{Anne Hutter$^{1,2,3}$\thanks{E-mail: anne.hutter@nbi.ku.dk}, Maxime Trebitsch$^1$, Pratika Dayal$^1$, Stefan Gottl\"ober $^4$, Gustavo Yepes $^{5,6}$,
\newauthor Laurent Legrand$^1$ \\
$^{1}$ Kapteyn Astronomical Institute, University of Groningen, P.O. Box 800, 9700 AV Groningen, The Netherlands \\
$^{{2}}$ Cosmic Dawn Center (DAWN) \\
$^{3}$ Niels Bohr Institute, University of Copenhagen, Jagtvej 128, DK-2200, Copenhagen N, Denmark \\
$^{4}$ Leibniz-Institut f\"ur Astrophysik, An der Sternwarte 16, 14482 Potsdam, Germany \\
$^{5}$ Departamento de Fısica Teorica, Modulo 8, Facultad de Ciencias, Universidad Autonoma de Madrid, 28049 Madrid, Spain\\
$^{6}$ CIAFF, Facultad de Ciencias, Universidad Autonoma de Madrid, 28049 Madrid, Spain
}

\date{Accepted 2023 July 18. Received 2023 July 18; in original form 2022 September 30}

\pubyear{2022}

\begin{document}
\label{firstpage}
\pagerange{\pageref{firstpage}--\pageref{lastpage}}
\maketitle

\begin{abstract}
We use the {\sc astraeus} framework to investigate how the visibility and spatial distribution of Lyman-$\alpha$ (Ly$\alpha$) emitters (LAEs) during reionisation is sensitive to a halo mass-dependent fraction of ionising radiation escaping from the galactic environment ($f_\mathrm{esc}$) and the ionisation topology. To this end, we consider the two physically plausible bracketing scenarios of $f_\mathrm{esc}$ increasing and decreasing with rising halo mass. We derive the corresponding observed Ly$\alpha$ luminosities of galaxies for three different analytic Ly$\alpha$ line profiles and associated Ly$\alpha$ escape fraction ($f_\mathrm{esc}^\mathrm{Ly\alpha}$) models: importantly, we introduce two novel analytic Ly$\alpha$ line profile models that describe the surrounding interstellar medium (ISM) as dusty gas clumps. They are based on parameterising results from radiative transfer simulations, with one of them relating $f_\mathrm{esc}^\mathrm{Ly\alpha}$ to $f_\mathrm{esc}$ by assuming the ISM of being interspersed with low-density tunnels. Our key findings are: (i) for dusty gas clumps, the Ly$\alpha$ line profile develops from a central to double peak dominated profile as a galaxy's halo mass increases; (ii) LAEs are galaxies with $M_h\gtrsim10^{10}\msun$ located in overdense and highly ionised regions; (iii) for this reason, the spatial distribution of LAEs is primarily sensitive to the global ionisation fraction and only weakly in second-order to the ionisation topology or a halo mass-dependent $f_\mathrm{esc}$; (iv) furthermore, as the observed Ly$\alpha$ luminosity functions reflect the Ly$\alpha$ emission from more massive galaxies, there is a degeneracy between the $f_\mathrm{esc}$-dependent intrinsic Ly$\alpha$ luminosity and the Ly$\alpha$ attenuation by dust in the ISM if $f_\mathrm{esc}$ does not exceed $\sim50\%$.
\end{abstract}

\begin{keywords}
galaxies: high-redshift - intergalactic medium - dark ages, reionisation, first stars - methods: numerical
\end{keywords}



\section{Introduction}
\label{sec_introduction}

The Epoch of Reionisation (EoR) marks the second major phase transition in the Universe. With the emergence of the first galaxies, ultraviolet (UV) radiation gradually ionises the neutral hydrogen (\HI) in the intergalactic medium (IGM) until the Universe is reionised by $z\simeq5.3$ \citep{Fan2006, Keating2020, Zhu2021, Bosman2022}. However, as only the brighter galaxies during the EoR are observed to date, key questions detailing the reionisation process remain outstanding: Did the few bright and more massive or the numerous faint and low-mass galaxies contribute more to reionisation? 
Feedback processes, such as heating by supernovae (SN) and photoionisation, suppress star formation in low-mass galaxies \citep{Gnedin2000b, Gnedin2014, ocvirk2016, Ocvirk2020, Hutter2021a}, and reduce the contribution of very low-mass galaxies to reionisation. An even more critical quantity that regulates the ionising radiation (with energies $E>13.6$~eV) escaping from galaxies and thus the galaxy population driving the reionisation of the IGM is the fraction of ionising photons $f_\mathrm{esc}$ that escape from galaxies into the IGM \citep[e.g.][]{kim2013b, Seiler2019, Hutter2021b, Garaldi2022}.

While the presence of \HI in the IGM during the EoR impedes direct measurements of $f_\mathrm{esc}$, different theoretical models and simulations have investigated the physical processes determining and dependencies of $f_\mathrm{esc}$ \citep[e.g.][]{Ferrara2013, Wise2014, Kimm2014, Kimm2019}. Cosmological radiation hydrodynamical simulations suggest that $f_\mathrm{esc}$ decreases towards deeper gravitational potential \citep[e.g.][]{Wise2014, Kimm2014, Kimm2017, Kimm2019, Xu2016, anderson2017, lewis2020}. High-resolution simulations of the ISM indicate that $f_\mathrm{esc}$ is dominated by the escape from star-forming clouds. The ionising radiation of massive stars and their explosions as SN ionise, heat and destroy the star-forming clouds clearing the way for the ionising radiation to escape \citep{Howard2018, Kim2019, He2020, Kimm2022}. The complex dependency of $f_\mathrm{esc}$ on the underlying gravitational potential, the gas distribution and stellar populations in the ISM leaves marks not only in the radiation emitted by galaxies but also in the ionisation topology, the time and spatial distribution of the ionised regions around galaxies.

Current and forthcoming observations of galaxies and the ionisation state of the IGM have the potential to constrain galactic properties, such as $f_\mathrm{esc}$, and the reionisation process. On the one hand, detecting the 21cm signal from \HI in the IGM with forthcoming large radio interferometers (e.g. Square Kilometre Array) will measure the ionisation topology, which provides constraints on the dependence of $f_\mathrm{esc}$ on galaxy mass \citep{kim2013b, Seiler2019, Hutter2020}.
On the other hand, being extremely sensitive to the attenuation by \HI in the IGM, the observable Lyman-$\alpha$ (Ly$\alpha$) radiation at $1216$\AA~ from high-redshift galaxies has gained popularity in probing reionisation for the following reason: A $z\gtrsim6$ galaxy only exhibits detectable Ly$\alpha$ emission when: (i) it is surrounded by an ionised region that is large and ionised (i.e. low residual \HI fraction) enough to allow a sufficient fraction of its emerging Ly$\alpha$ line to traverse the IGM, or (ii) it is gas-rich enough (corresponding to a high \HI column density) such that the red part of the Ly$\alpha$ line emerging from the galaxy is redshifted out of absorption, or (iii) it has strong outflows that redshift the emerging Ly$\alpha$ line out of absorption, or it is a combination of all three. The first criterion suggest that more massive galaxies able to retain more gas might be the most likely to show observable Ly$\alpha$ emission during the EoR: their higher rates of forming stars emitting ionising photons lead to an increased production of Ly$\alpha$ radiation in the ISM and the growth of large ionised regions around them. The latter is accelerated by their ionised regions merging earlier with those of the surrounding lower mass objects attracted by their deeper gravitational potentials \citep{Chardin2012, Furlanetto2016, Chen2019}. As reionisation progresses and the ionised regions grow, increasingly lower mass galaxies become visible as Ly$\alpha$ emitters (LAEs), which leads not only to a higher fraction of galaxies showing Ly$\alpha$ emission but also to a reduced clustering of LAEs \citep{mcquinn2007, jensen2013, Hutter2015, Sobacchi2015}.

This picture is increasingly supported by observations of $z>6$ LAEs. Not only the fraction of Lyman Break Galaxies (LBGs) showing Ly$\alpha$ emission rises from $z\simeq8$ to $z\simeq6$ \citep{Schenker2014, Pentericci2014, Pentericci2018, Fuller2020}, but also the majority of Ly$\alpha$ emission at $z\gtrsim6.5$ is detected in galaxies with a bright UV continuum \citep{Oesch2015, Zitrin2015, Roberts-Borsani2016, Endsley2022, Endsley_Stark2022}. 
Moreover, the close proximity of UV-bright LAEs suggests that LAEs are located in over-dense regions \citep{Vanzella2011, Castellano2016, Castellano2018, Jung2020, Tilvi2020, Hu2021, Endsley_Stark2022} that exhibit the first and largest ionised regions during the EoR. This hypothesis is also in line with the observed double-peaked Ly$\alpha$ profiles in $z\gtrsim6.5$ galaxies \citep{Songaila2018, Hu2016, Matthee2018, Meyer2021}, indicating that the ionised regions surrounding them are so large that even the part bluewards the Ly$\alpha$ resonance redshifts out of resonance. 
Current theoretical predictions of the large-scale LAE distribution confirm this picture, suggesting that the LAEs we see during the EoR are more massive galaxies naturally located in over-dense regions \citep[c.f.][]{Dayal2011, jensen2013, Hutter2014, Mesinger2015, Weinberger2018, Qin2022}. 

Yet, all these LAE models effectively assume a constant $f_\mathrm{esc}$ value across the entire galaxy population at a given redshift. This assumption remains highly uncertain as $f_\mathrm{esc}$ is very sensitive to the ISM and the circumgalactic medium (CGM) of galaxies that again depend on the underlying gravitational potential of a galaxy. However, it is essential, since $f_\mathrm{esc}$ defines the critical processes that shape the Ly$\alpha$ luminosities observed from galaxies. An $f_\mathrm{esc}$ varying with galactic properties and the underlying gravitational potential might alter the galaxy population seen as LAEs for the following reasons:
Firstly, within a galaxy, most Ly$\alpha$ radiation is produced by recombining hydrogen atoms (see e.g. \citet{Laursen2019} and \citet{Faucher-Giguere2010} for an estimate showing that Ly$\alpha$ cooling radiation is subdominant) and scales with the number of \HI ionising photons absorbed within the galaxy ($\propto 1-f_\mathrm{esc}$). Secondly, a fraction of these Ly$\alpha$ photons undergoes only a few scattering events when they escape through the same low-density tunnels that facilitate the escape of \HI ionising photons. In contrast, the other fraction that traverses optically thick clouds upon its escape is scattered and absorbed by hydrogen and dust, respectively \citep[see, e.g.][]{Verhamme2015, Dijkstra2016, Kimm2019, Kakiichi2021}. These different escape mechanisms result not only in $f_\mathrm{esc}$ posing a lower limit to the fraction of Ly$\alpha$ photons escaping from a galaxy but also determining the Ly$\alpha$ line profile that emerges from a galaxy. Detailed low-redshift galaxy observations increasingly supported the $f_\mathrm{esc}$-sensitivity of these Ly$\alpha$ properties \citep{Verhamme2017, Jaskot2019, Gazagnes2020}. Thirdly, $f_\mathrm{esc}$ shapes the IGM ionisation topology by determining the number of ionising photons available to ionise the IGM surrounding a galaxy. 
While a higher $f_\mathrm{esc}$ value enlarges the ionised region surrounding a galaxy and enhances the transmission of Ly$\alpha$ radiation through the IGM \citep{Dayal2011, Hutter2014}, the corresponding Ly$\alpha$ line emerging from a galaxy will be more peaked around the Ly$\alpha$ resonance and raise the absorption by \HI in the IGM.
Given this complex $f_\mathrm{esc}$-dependency of the observed Ly$\alpha$ luminosity, it remains unclear whether different dependencies of $f_\mathrm{esc}$ with galaxy properties (e.g. increasing or decreasing with rising halo mass) would (i) identify the same galaxies as LAEs (exceeding a threshold Ly$\alpha$ luminosity) and/or (ii) lead to different spatial large-scale distribution of the LAEs' Ly$\alpha$ luminosities. In other words, which of these $f_\mathrm{esc}$-dependent Ly$\alpha$ processes dominates the observed Ly$\alpha$ luminosities? For example, is the $f_\mathrm{esc}$-dependency of the intrinsic Ly$\alpha$ luminosity dominant, and we yield a weaker clustering of LAEs when $f_\mathrm{esc}$ value decreases with rising halo mass? Or do they compensate each other once we reproduce the observed Ly$\alpha$ luminosity functions (Ly$\alpha$ LFs)? 

To address these questions, we use our {\sc astraeus} framework that models galaxy evolution and reionisation self-consistently \citep{Hutter2021a, Ucci2023}, and simulate different reionisation scenarios that gauge the physically plausible range of $f_\mathrm{esc}$ dependencies, i.e. $f_\mathrm{esc}$ decreasing and increasing with rising halo mass. Moreover, we parameterise results from numerical Ly$\alpha$ radiative transfer (RT) simulations of clumpy media \citep{Gronke2017} and build an analytic model for the fraction of Ly$\alpha$ photons escaping and the corresponding Ly$\alpha$ line profile emerging from high-redshift galaxies. Importantly, we explore three different Ly$\alpha$ line profile models, including (i) a Gaussian profile around the Ly$\alpha$ resonance where the Ly$\alpha$ escape fraction is directly related to the dust attenuation of the UV continuum \citep[used in previous LAE models outlined in][]{Dayal2011, Hutter2014}, (ii) a Ly$\alpha$ line profile emerging from a shell of dusty gas clumps, which we model by using the different Ly$\alpha$ escape regimes identified in \citet{Gronke2017}, and (iii) a Ly$\alpha$ line profile emerging from a shell of gas clumps with a fraction $f_\mathrm{esc}$ of the solid angle interspersed by gas-free tunnels. The latter two give rise to various combinations of a central peak around the Ly$\alpha$ resonance (Ly$\alpha$ photons hardly scatter in an optically thin medium) and two peaks in the red and blue wings (Ly$\alpha$ photons are scattered in an optically thick medium). By deriving the observed Ly$\alpha$ luminosities of all simulated galaxies for all combinations of reionisation scenarios and Ly$\alpha$ line models, we address the following questions: Which $f_\mathrm{esc}$-dependent Ly$\alpha$ process, i.e. intrinsic production, escape or transmission through the IGM of Ly$\alpha$ radiation, dominates the observed Ly$\alpha$ luminosity? Can the observed Ly$\alpha$ luminosities of galaxies inform us on their emerging Ly$\alpha$ line profile?
Given the ionisation topology depends sensitively on the assumed dependency of $f_\mathrm{esc}$ with halo mass, are the same or different galaxies identified as LAEs and do they differ in the spatial distribution of their Ly$\alpha$ luminosities?

This paper is organised as follows. In Section \ref{sec_model} we briefly describe the {\sc astraeus} model, its implementation of dust and the different reionisation simulations. In Section \ref{sec_modelling_LAEs} we introduce the different Ly$\alpha$ line profile models and their corresponding attenuation by dust. We then (Section \ref{sec_number_and_properties_LAEs}) discuss how the Ly$\alpha$ line profiles depend on halo mass in our different reionisation scenarios, how free model parameters, such as the ISM clumpiness or size of the dust gas clumps, need to be adjusted to fit the observed Ly$\alpha$ LFs, and how the galaxy properties determining the observed Ly$\alpha$ luminosities depend on the halo mass of a galaxy. In Section \ref{sec_spatial_distribution_LAEs} we identify the location of LAEs in the large-scale density and ionisation structure and assess whether the spatial distribution of LAEs differs for different $f_\mathrm{esc}$-dependencies on halo mass/ionisation topologies. Finally, we briefly discuss which Lyman Break galaxies are preferentially identified as LAEs (Section \ref{sec_LAE_LBG_relation}) and conclude in Section \ref{sec_conclusions}.
In this paper we assume a $\Lambda$CDM Universe with cosmological parameter values of $\Omega_\Lambda=0.69$, $\Omega_m=0.31$, $\Omega_b=0.048$, $H_0=100h=67.8$km~s$^{-1}$Mpc$^{-1}$, $n_s=0.96$ and $\sigma_8=0.83$, and a Salpeter initial mass function \citep[IMF;][]{salpeter1955} between $0.1\msun$ to $100\msun$.

\section{The model and simulations}
\label{sec_model}

In this paper, we use the {\sc astraeus} framework. This framework couples a semi-analytic galaxy evolution model \citep[an enhanced version of {\sc delphi};][]{Dayal2014} with a semi-numerical reionisation scheme \citep[{\sc cifog};][]{Hutter2018a} and runs the resulting model on the outputs of a dark matter (DM) only N-body simulation. In this Section, we provide a brief description of the physical processes implemented in {\sc astraeus} \citep[for more details, see][]{Hutter2021a} and introduce the different reionisation simulations. 

\subsection{N-body simulation}
\label{subsec_Nbody_simulation}

As part of the Multidark simulation project, the underlying DM N-body simulation ({\sc very small multidark planck; vsmdpl}) has been run with the {\sc gadget-2 tree+pm} code \citep{springel2005}. In a box with a side length of $160h^{-1}$Mpc, it follows the trajectories of $3840^3$ DM particles. Each DM particle has a mass of $6\times10^6 h^{-1}\msun$. For a total of $150$ snapshots ranging from $z=25$ to $z=0$, the phase space {\sc rockstar} halo finder \citep{behroozi2013_rs} has been used to identify all halos and subhalos down to $20$ particles or a minimum halo mass of $1.24 \times 10^8h^{-1}\msun$. To obtain the local horizontal merger trees (sorted on a redshift-by-redshift basis within a tree) for galaxies at $z=4.5$ that {\sc astraeus} requires as input, we have used the pipeline internal {\sc cutnresort} scheme to cut and resort the vertical merger trees (sorted on a tree-branch by tree-branch basis within a tree) generated by {\sc consistent trees} \citep{behroozi2013_trees}. For the first $74$ snapshots that range from $z=25$ to $z=4.5$, we have generated the DM density fields by mapping the DM particles onto $2048^3$ grids and re-sampling these to $512^3$ grids used as input for the {\sc astraeus} pipeline.

\subsection{Galaxy evolution}
\label{subsec_galaxy_evolution}

{\sc astraeus} tracks key processes of early galaxy formation and reionisation by post-processing the DM merger trees extracted from the {\sc vsmdpl} simulation. At each time step (i.e. snapshot of the N-body simulation) and for each galaxy, it tracks the amount of gas that is accreted, the gas and stellar mass merging, star formation and associated feedback from SNII and metal enrichment, as well as the large-scale reionisation process and its associated feedback on the gas content of early galaxies.

\subsubsection{Gas and stars}
\label{subsubsec_gas_and_stars}

In the beginning, when a galaxy starts forming stars in a halo with mass $M_h$, it has a gas mass of $M_g^i(z)=f_g (\Omega_b/\Omega_m) M_h(z)$, with $f_g$ being the gas fraction not evaporated by reionisation, i.e. $f_g=1$ and $f_g<1$ as the galaxy forms in a neutral and ionised region, respectively. 
In subsequent time steps a galaxy gains gas from its progenitors ($M_g^\mathrm{mer}(z)$) and smooth accretion ($M_g^\mathrm{acc}$), while its total gas mass never exceeds the limit given by reionisation feedback: 
\begin{eqnarray}
M_g^i(z) &=& \min\left( M_g^\mathrm{mer}(z) + M_g^\mathrm{acc}(z), f_g (\Omega_m/\Omega_b) M_h \right)
\end{eqnarray}
with 
\begin{eqnarray}
M_g^\mathrm{acc}(z) &=& M_h(z) - \sum_{p=1}^\mathrm{N_p} M_{h,p}(z + \Delta z) \\
M_g^\mathrm{mer}(z) &=& \sum_{p=1}^\mathrm{N_p} M_{h,p} (z + \Delta z),
\end{eqnarray}
where $N_p$ is the galaxy's number of progenitors and $M_{h,p}$ the halo mass of each progenitor.

At each time step, a fraction of the merged and accreted (initial) gas mass is transformed into stellar mass, $M_\star^\mathrm{new}(z)=(f_\star^\mathrm{eff}/\Delta t) M_g^i(z)$.\footnote{We note that this definition has been altered compared to the first version of {\sc astraeus} in \citep{Hutter2021a}.} Here $f_\star^\mathrm{eff}$ represents the fraction of gas that forms stars over a time span $\Delta t$ and is limited by the minimum amount of stars that need to form to eject all gas from the galaxy, $f_\star^\mathrm{ej}$, and an upper limit, $f_\star$. $f_\star^\mathrm{eff}$ depends on the gravitational potential: more massive galaxies form stars at the constant rate $f_\star$, while low-mass galaxies form stars at the limited rate $f_\star^\mathrm{ej}$ due to SN and radiative feedback. While we account for radiative feedback from reionisation by modifying the initial gas mass reservoir with the factor $f_g$, $f_\star^\mathrm{eff}$ incorporates the suppression of star formation in low-mass halos as gas is heated and ejected by SNII explosions. Our model incorporates a delayed SN feedback scheme, i.e. at each time step the effective star formation efficiency accounts for the SNII energy released from stars formed in the current and previous time steps, following the mass-dependent stellar lifetimes \citep{padovani1993}. In contrast to \citet{Hutter2021a}, we have updated our model and do not assume stars to form in bursts to calculate the number of SNII exploding within a time step but $M_\star^\mathrm{new}(z)$ to form at a constant star formation over the entire time step (see Appendix \ref{app_delayed_non-bursty_SNscheme} for a detailed calculation). The star formation efficiency in the SN feedback-limited regime is given by 
\begin{eqnarray}
f_\star^\mathrm{ej}(z) &=& \frac{v_c^2}{v_c^2 + f_w E_{51} \nu_z} \left[  1 - \frac{f_w E_{51} \sum_j \nu_j M_{\star,j}^\mathrm{new}(z_j)}{M_\mathrm{g}^i(z)~ v_c^2} \right],
\end{eqnarray}
with $v_c$ being the rotational velocity of the halo, $E_{51}$ the energy released by a SNII, $f_w$ the fraction of SNII energy injected into the winds driving gas outflows, $M_{\star,j}^\mathrm{new}(z_j)$ the stellar mass formed during previous time steps $j$, and $\nu_j$ the fraction of stellar mass formed in previous time step $j$ that explodes in the current time step given the assumed IMF. 

{\sc astraeus} incorporates multiple models for radiative feedback from reionisation, ranging from a weak and time-delayed ({\it Weak Heating}) to a strong instantaneous feedback ({\it Jeans mass}). In this work, we use the intermediate and time-delayed {\it Photoionisation} model, where the characteristic mass defining the gas fraction not evaporated by reionisation grows on a dynamical timescale to the respective Jeans mass \citep[for a detailed description see][]{Hutter2021a}. We list the {\sc astraeus} model parameters and their assumed values in Table \ref{table_model_params}. $f_\star$ and $f_w$ have been adjusted to reproduce the observed UV LFs, stellar mass functions, global star formation rate density, and global stellar mass density at $z=10-5$.

\begin{table}
  \centering
  \caption{{\sc astraeus} model parameters and chosen values in this work.}
  \label{tab:example_table}
  \begin{tabular*}{\columnwidth}{ccc}
    \hline\hline
    Parameter & Value or reference & Description\\
    \hline
    $f_\star$ & $0.025$ & Maximum star-formation efficiency \\
    $f_w$      & $0.2$   & SN coupling efficiency \\
    - & Photoionization & Radiative feedback model \\
    IMF & \citet{salpeter1955} & For stellar evolution, enrichment, SED \\
    SED & \textsc{Starburst99} & ionizing SED model \\
    \hline
  \end{tabular*}
  \label{table_model_params}
\end{table}

\subsubsection{Metals and dust}
\label{subsubsec_metals_and_dust}

The current {\sc astraeus} model also incorporates the metal enrichment by stellar winds, SNII and SNIa explosions \citep[for a detailed description see][]{Ucci2023}. At each time step, we assume that gas smoothly accreted has the average metallicity of the gas in the IGM, $Z_\mathrm{IGM}$. Metals are produced through stellar winds, SNII and SNIa explosions. The amount of newly forming metals depends on the number of massive stars exploding as SN in the current time step according to \citet{padovani1993}, \citet{yates2013} and \citet{maoz2012}. For the corresponding stellar metal yields, {\sc astraeus} uses the latest yield tables from \citet{Kobayashi2020b}. We assume that gas and metals are perfectly mixed. Thus, the metals ejected from the galaxy are proportional to the ejected gas mass and the metallicity of the gas in the galaxy. This ejected metal mass contributes to $Z_\mathrm{IGM}$.

In this work, we have extended the {\sc astraeus} model \citep{Hutter2021a, Ucci2023} to follow the formation, growth, destruction, astration and destruction of dust in each galaxy \citep[c.f.][for details]{Dayal2022}. We note that we consider dust to be part of our metal reservoir (i.e. $M_\mathrm{dust}\leq M_\mathrm{m}$). At each time step, {\sc astraeus} computes the evolution of the dust mass $M_\mathrm{dust}$ in a galaxy by solving the following differential equation
\begin{eqnarray}
    \frac{\mathrm{d}M_\mathrm{dust}}{\mathrm{d}t} &=& \dot{M}_\mathrm{dust}^\mathrm{prod} + \dot{M}_\mathrm{dust}^\mathrm{grow} - \dot{M}_\mathrm{dust}^\mathrm{dest} - \dot{M}_\mathrm{dust}^\mathrm{astr} - \dot{M}_\mathrm{dust}^\mathrm{ej}.
\label{eq_dust}
\end{eqnarray}
The first term on the right hand side (RHS) of Eqn. \ref{eq_dust} denotes the production of dust in SNII and AGB stars through condensation of metals in stellar ejecta
\begin{eqnarray}
    \dot{M}_\mathrm{dust}^\mathrm{prod} &=& y_\mathrm{SNII} \gamma_\mathrm{SNII} + \dot{M}_\mathrm{dust}^\mathrm{AGB},
\end{eqnarray}
with $y_\mathrm{SNII}=0.45\msun$ being the dust mass formed per SNII, 
\begin{eqnarray}
    \gamma_\mathrm{SN}(t) &=& \int_{8\msun}^{40\msun} \mathrm{SFR}(t - \tau_m) \phi(m) \mathrm{d}m
\end{eqnarray}
the number of SNII events,
\begin{eqnarray}
    \dot{M}_\mathrm{dust}^\mathrm{AGB}(t) &=& \int_{0.85\msun}^{50\msun} y_\mathrm{AGB}(m) \mathrm{SFR}(t - \tau_m) \phi(m) \mathrm{d}m
\end{eqnarray}
the contribution from AGB stars and $y_\mathrm{AGB}$ the dust yields from AGB stars. In agreement with \citep{Ucci2023}, we adopt the latest yield tables from \citet{Kobayashi2020b} for $y_\mathrm{AGB}$.
The second term on the RHS of Eqn. \ref{eq_dust} describes the dust grain growth through the accretion of heavy elements in dense molecular clouds in the ISM,
\begin{eqnarray}
    \dot{M}_\mathrm{dust}^\mathrm{grow} &=& \left( Z' - \frac{M_\mathrm{dust}}{M_\mathrm{g}^i} \right) f_\mathrm{cold~gas} \frac{M_\mathrm{dust}}{\tau_\mathrm{gg,0} Z_\odot}
\end{eqnarray}
where $Z'$ is the metallicity after accretion and star formation, $M_\mathrm{dust}$ is the dust mass, $f_\mathrm{cold~gas}$ the fraction of cold and molecular gas, and $\tau_\mathrm{gg}=\tau_\mathrm{0,gg} / Z$ the accretion timescale adopted from \citet{Asano2013} \citep[see also][]{Triani2020}. We assume $f_\mathrm{cold~gas}=0.5$ and $\tau_\mathrm{gg,0}=30$Myrs.
The third term in Eqn. \ref{eq_dust} describes the destruction of dust by SN blastwaves, for which we adopt the analytic description outlined in \citet{mckee1989}
\begin{eqnarray}
    \dot{M}_\mathrm{dust}^\mathrm{dest} &=& \left( 1 - f_\mathrm{cold~gas} \right) \frac{M_\mathrm{dust}}{M_\mathrm{g}^i}\ \gamma_\mathrm{SN} \epsilon\ M_\mathrm{SN, bw},
\end{eqnarray}
with $\epsilon$ being the effifiency of dust destruction in a SN-shocked ISM and $M_\mathrm{SN, bw}$ the mass accelerated to $100$~km~s$^{-1}$ by the SN blast wave. In line with \citet{mckee1989} and \citet{lisenfeld_ferrara1998} we adopt $\epsilon=0.03$ and $M_\mathrm{SN, bw}=6.8\times10^3\msun$.
Finally, Eqn. \ref{eq_dust} accounts also for the destruction of dust by astration as new stars form from the metal-enriched gas,
\begin{eqnarray}
    \dot{M}_\mathrm{dust}^\mathrm{astr} &=& Z^\mathrm{i}\ \frac{M_\star^\mathrm{new}}{\Delta t},
\end{eqnarray}
and the ejection of metals through winds powered by the energy injected by SN,
\begin{eqnarray}
    \dot{M}_\mathrm{dust}^\mathrm{ej} &=& Z'\ \frac{M_\mathrm{g}^\mathrm{ej}}{\Delta t}.
\end{eqnarray}
The parameter values ($y_\mathrm{SNII}$, $\tau_\mathrm{gg,0}$, $\epsilon$, $M_\mathrm{SN,bw}$) quoted reasonably reproduce the observed UV LFs when the UV is attenuated by dust as follows (please see \citet{Hutter2021a} for observational UV LFs data points included): From the dust mass, $M_d$, we obtain the total optical depth to UV continuum photons as \citep[see e.g.][]{Dayal2011}
\begin{eqnarray}
    \tau_\mathrm{UV,c} &=& \frac{3 \Sigma_d}{4as},
\end{eqnarray}
with $\Sigma = M_d / (\pi r_d^2)$ being the dust surface mass density, $r_d$ the dust distribution radius, and $a=0.03~\mu$m and $s=2.25$g~cm$^{-3}$ the radius and material density of graphite/carbonaceous grains \citep{todini-ferrara2001}. Since we assume that dust and gas are perfectly mixed, we equate the dust distribution radius, $r_d$, with the radius of the gas, $r_g=4.5 \lambda r_\mathrm{vir} \left[ (1+z)/6 \right]^{1.8}$. Here $\lambda$ is the spin parameter of the simulated halo, $r_\mathrm{vir}$ the virial radius, and the third factor accounts for the redshift evolution of the compactness of galaxies and ensures that the observed UV LFs at $z=5-10$ are well reproduced. 
For a slab-like geometry, the escape fraction of UV continuum photons of a galaxy is then given by
\begin{eqnarray}
    f_\mathrm{esc}^\mathrm{c} &=& \frac{1-\exp(-\tau_\mathrm{UV,c})}{\tau_\mathrm{UV,c}},
\end{eqnarray}
and its observed UV luminosity by
\begin{eqnarray}
    L_\mathrm{c}^\mathrm{obs} &=& f_\mathrm{esc}^\mathrm{c} L_\mathrm{c},
\end{eqnarray}
with the intrinsic UV luminosity, $L_\mathrm{c}$, being computed as outlined in Section 2.2.4 in \citet{Hutter2021a}.

\subsection{Reionisation}
\label{subsec_reionisation}

At each time step {\sc astraeus} follows the time and spatial evolution of the ionised regions in the IGM. For this purpose, it derives the number of ionising photons produced in each galaxy, $\dot{Q}$, by convolving the galaxy's star formation rate history with the spectra of a metal-poor ($Z=0.05$Z$_\odot$) stellar population. Spectra have been obtained from the stellar population synthesis model {\sc starburst99} \citep{Leitherer1999}. Again we assume that stars form continuously over a time step. Then the number of ionising photons that contribute to the ionisation of the IGM is then given by
\begin{eqnarray}
    \dot{N}_\mathrm{ion} &=& f_\mathrm{esc} \dot{Q},
\end{eqnarray}
where $f_\mathrm{esc}$ is the fraction of ionising photons that escape from the galaxy into the IGM. From the resulting ionising emissivity and gas density distributions {\sc astraeus} derives the spatial distribution of the ionised regions by comparing the cumulative number of ionising photons with the number of absorption events \citep[see {\sc cifog},][, for details]{Hutter2018a}. Within ionised regions, it also derives the photoionisation rate and residual \HI fraction in each grid cell. The ionisation and photoionisation fields obtained allow us then to determine on the fly whether the environment of a galaxy has been reionised and account for the corresponding radiative feedback by computing the gas mass the galaxy can hold on to ($f_g M_g^i$).

\subsection{Simulations}
\label{subsec_simulations}

\begin{figure}
    \centering
    \includegraphics[width=0.5\textwidth]{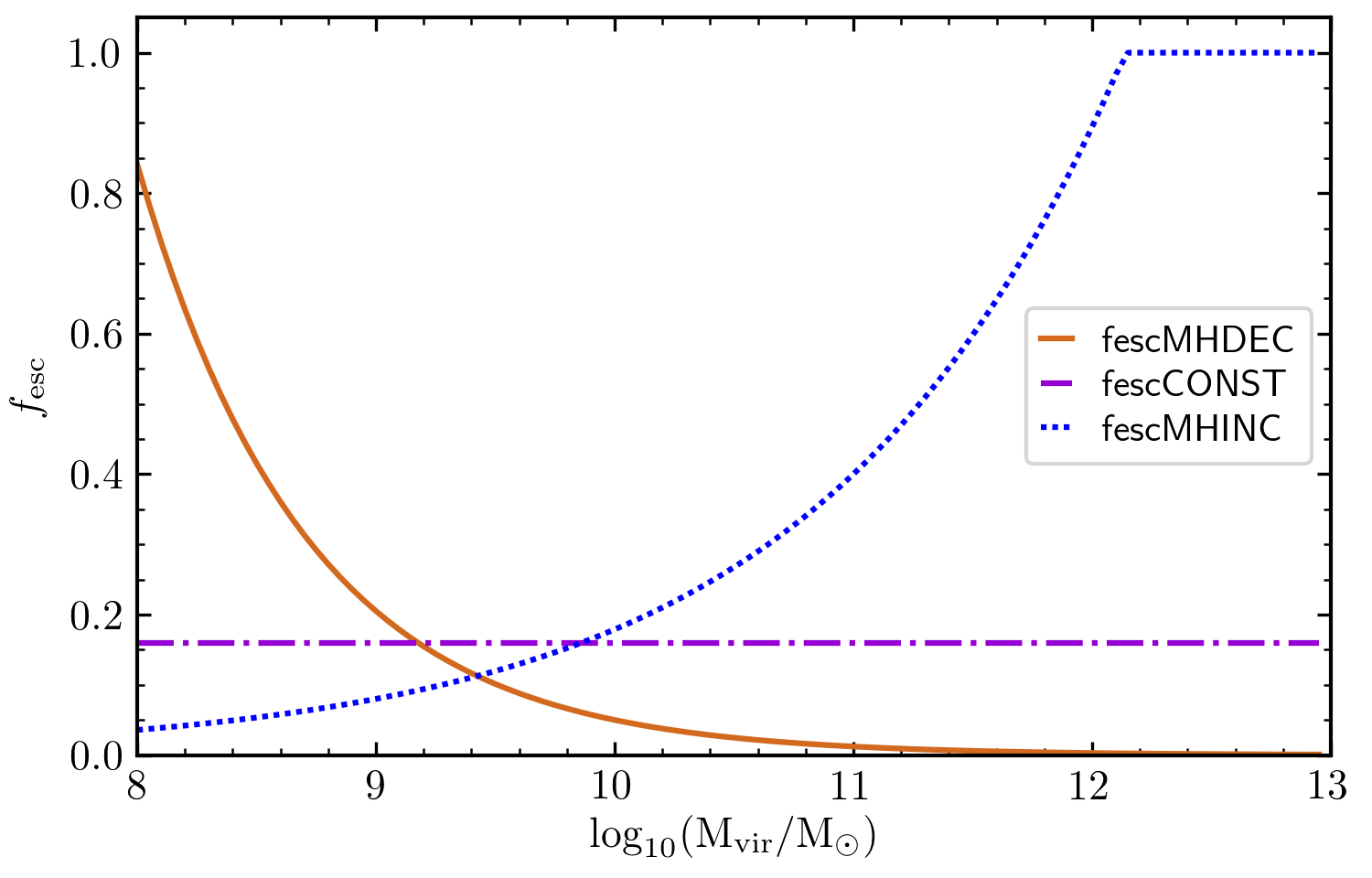}
    \caption{The ionising escape fraction $f_\mathrm{esc}$ for the three models, decreasing (solid orange line), being constant (dash dotted magenta line) and increasing (dotted blue line) with halo mass $M_h$.}
    \label{fig_fesc}
\end{figure}
\begin{figure}
    \centering
    \includegraphics[width=0.5\textwidth]{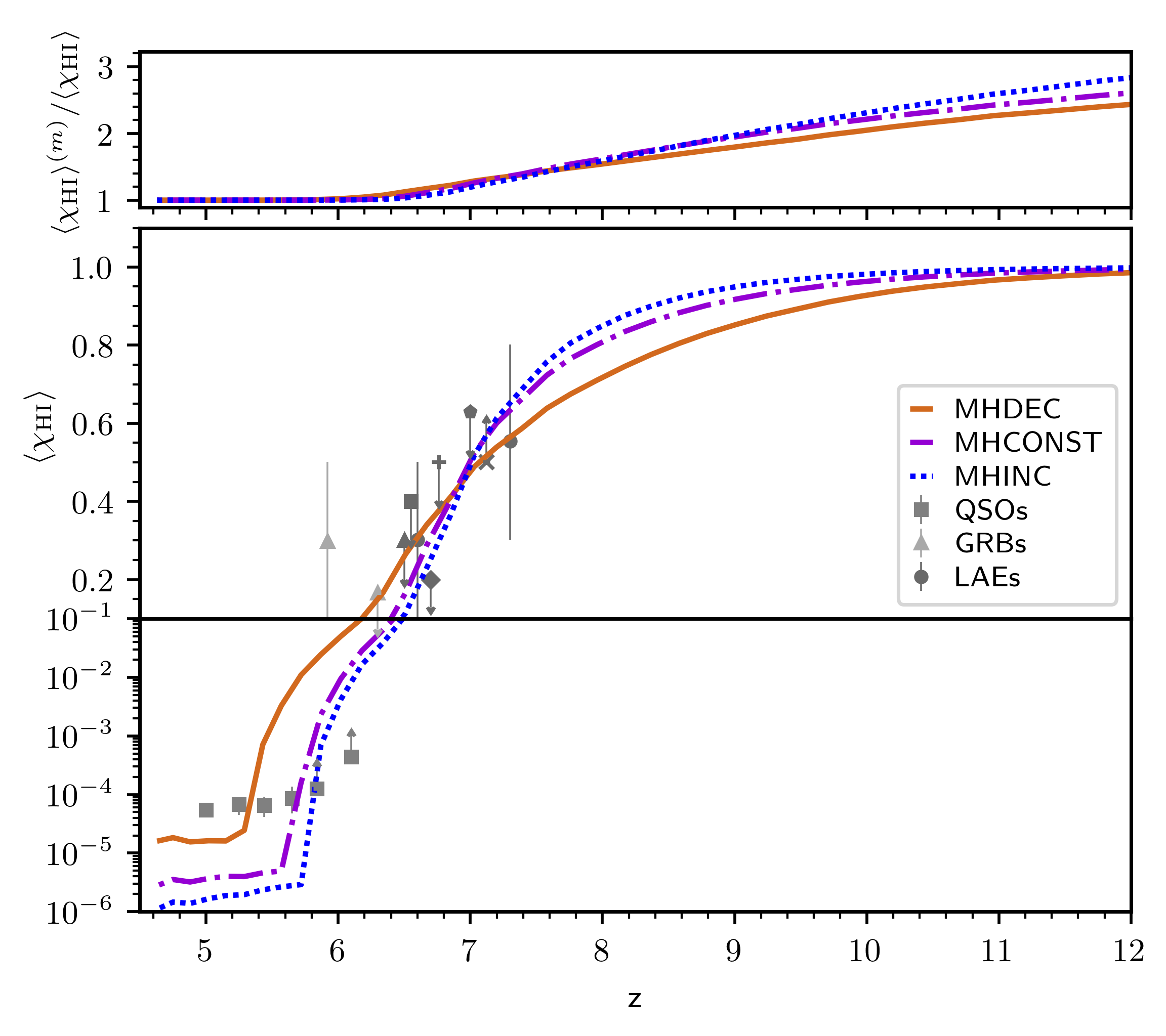}
    \caption{Ratio of the mass- and volume-averaged neutral hydrogen fraction (top panel) and volume averaged neutral hydrogen fraction (bottom panel) as a function of redshift. In each panel, we show results for our three $f_\mathrm{esc}$ models: decreasing (solid orange line), being constant (dash dotted magenta line) and increasing (dotted blue line) with halo mass $M_h$. In the lower panel, grey points indicate observational constraints from: GRB optical afterglow spectrum analyses \citep[light triangles;][]{Totani2006, Totani2014}, quasar sightlines \citep[Medium squares;][]{Fan2006}, Lyman-$\alpha$ LFs  \citep[dark circles]{Konno2018}, \citep[dark squares;][]{Kashikawa2011}, \citep[dark diamonds][]{Ouchi2010}, \citep[dark pentagons][]{Ota2010} and \citep[dark triangles][]{Malhotra2004}, Lyman-$\alpha$ emitter clustering \citep[dark plus signs;][]{Ouchi2010} and the Lyman-$\alpha$ emitting galaxy fraction \citep[dark crosses;][]{Pentericci2011, Schenker2012, Ono2012, Treu2012, Caruana2012, Caruana2014, Pentericci2014}.}
    \label{fig_hist_ion}
\end{figure}

In the following we consider three different reionisation scenarios that explore the physically plausible space of the ionising escape fraction $f_\mathrm{esc}$ (c.f. Fig. \ref{fig_fesc}):
\begin{enumerate}
    \item {\sc mhdec}: $f_\mathrm{esc}$ decreases with rising halo mass of a galaxy (red solid line)
    \begin{eqnarray}
        f_\mathrm{esc} &=& f_\mathrm{esc,low} \left( \frac{f_\mathrm{esc,high}}{f_\mathrm{esc,low}} \right)^{\frac{\log_{10} (M_h / M_{h,\mathrm{low}}) }{\log_{10} (M_{h,\mathrm{high}}  / M_{h,\mathrm{low}} )}}
    \label{eq_fesc}
    \end{eqnarray} 
    with $f_\mathrm{esc,low}=0.55$, $f_\mathrm{esc,high}=0.05$, $M_{h,\mathrm{low}}=2\times10^8h^{-1}\msun$ and $M_{h,\mathrm{high}}=10^{10}h^{-1}\msun$.
    \item {\sc mhconst}: $f_\mathrm{esc}=0.16$ for each galaxy (magenta dash-dotted line).
    \item {\sc mhinc}: $f_\mathrm{esc}$ increases with rising halo mass of a galaxy (blue dotted line) following Eqn. \ref{eq_fesc} with $f_\mathrm{esc,low}=0.08$, $f_\mathrm{esc,high}=0.4$, $M_{h,\mathrm{low}}=10^9h^{-1}\msun$ and $M_{h,\mathrm{high}}=10^{11}h^{-1}\msun$.
\end{enumerate}
These three $f_\mathrm{esc}$ prescriptions have been adjusted to reproduce the electron optical depth measured by Planck \citep{planck2018} and fit the observational constraints from LAEs, quasar absorption spectra and gamma ray bursts (as depicted in the lower panel of Fig. \ref{fig_hist_ion}). In addition, for {\sc mhinc} the maximum $f_\mathrm{esc}$ value of more massive galaxies is also limited by the observed Ly$\alpha$ LFs. Despite having very similar electron optical depths, these three $f_\mathrm{esc}$ prescriptions lead to different ionisation histories and topologies (see Fig. \ref{fig_hist_ion} and \ref{fig_XHImaps_with_LAEs}). As $f_\mathrm{esc}$ decreases with rising halo mass, reionisation is dominated by the low-mass galaxies ($M_h\lesssim10^{10}\msun$), leading to on average smaller ionised regions and lower photoionisation rates. Since these low-mass galaxies appear earlier, reionisation begins earlier (see solid red line in Fig. \ref{fig_hist_ion}); however, as shown in \citet{Hutter2021a} for the {\it Photoionisation} model their overall star formation rate decreases around $z\simeq7$, resulting in the Universe being reionised at a later time and exhibiting a higher average residual \HI fraction in ionised regions. In contrast, as $f_\mathrm{esc}$ increases with rising halo mass, more massive galaxies ($M_h\gtrsim10^{10}\msun$) drive reionisation. On average, ionised regions are larger and more clustered around more massive galaxies, and photoionisation rates within these ionised regions are higher. Reionisation begins later with the appearance of more massive galaxies and ends earlier as the abundance of these massive galaxies increases.

\section{Modelling Ly$\alpha$ emitters}
\label{sec_modelling_LAEs}

In this Section, we introduce the different models for the emergent Ly$\alpha$ line profiles (Section \ref{subsec_Lya_line_models}) and fractions of Ly$\alpha$ radiation escaping from a galaxy (Section \ref{subsec_dust_attenuation}), describe the attenuation of Ly$\alpha$ radiation by \HI in the IGM, and the derivation of the observed Ly$\alpha$ luminosity of a galaxy (Section \ref{subsec_IGM_attenuation}). We summarise the combinations of emerging Ly$\alpha$ line profile and dust attenuation models investigated in this paper in Section \ref{subsubsec_emerging_Lya_profile_models}.

\subsection{Emerging Ly$\alpha$ line profiles}
\label{subsec_Lya_line_models}

We investigate three Ly$\alpha$ line profiles $J(x)$: (1) a thermally Doppler-broadened Gaussian centred at the Ly$\alpha$ resonance; (2) a single, double or triple-peaked profile that depends on the clumpiness and \HI column density of the gas in a galaxy; (3) a single, double or triple-peaked profile that depends both on the ionising escape fraction $f_\mathrm{esc}$ and the clumpiness and \HI column density of the gas in a galaxy. While the first model represents a simple assumption used in previous works \citep[e.g.][]{Dayal2011, Hutter2014}, the latter two models are inspired by observations and detailed Ly$\alpha$ radiative transfer simulations \citep[e.g.][]{Dijkstra2016, Gronke2017}.
The Ly$\alpha$ line emerging from a galaxy is given by the intrinsic Ly$\alpha$ luminosity, $L_\alpha^\mathrm{intr}= \frac{2}{3}Q(1-f_\mathrm{esc})~h\nu_\alpha$, the escape fraction Ly$\alpha$ photons from the galaxy, $f_\mathrm{esc}^\mathrm{Ly\alpha}$, and the line profile $J(x)$.
\begin{eqnarray}
    L_\alpha^\mathrm{gal}(x) &=& L_\alpha^\mathrm{intr} f_\mathrm{esc}^\mathrm{Ly\alpha} J(x)
\end{eqnarray}
Here we have expressed the frequency deviation from the Ly$\alpha$ resonance $\nu_\alpha$ in terms of the thermal line broadening $\sigma_\mathrm{th}=(v_\mathrm{th}/c)\nu_\alpha$ with $v_\mathrm{th}=\sqrt{2k_B T/m_\mathrm{H}}$, yielding $x=\frac{\nu - \nu_\alpha}{\sigma_\mathrm{th}}$. $k_B$ is the Boltzmann constant, $m_\mathrm{H}$ the mass of a hydrogen atom and $T$ the temperature of the \HI gas.
In the remainder of this Section, we detail our different models for the Ly$\alpha$ line profiles and escape fractions.

\subsubsection{Central Gaussian}
\label{subsubsec_Lya_line_model_gaussian}

This model assumes that the emission sites of Ly$\alpha$ radiation, the hydrogen atoms within a galaxy, move at velocities that reflect the galaxy's rotation. The corresponding Doppler-broadened Ly$\alpha$ line profile is then given by
\begin{eqnarray}
    J_\mathrm{centre}(x) &=& \frac{1}{\sqrt{\pi}} \frac{\sigma_\mathrm{th}}{\sigma_r} \exp \left[ -x^2 \frac{\sigma_\mathrm{th}^2}{\sigma_r^2} \right],
\label{eq_Jcentre}
\end{eqnarray}
We note that since the $\sigma_\mathrm{th}$-dependence of $x$ cancels any dependency of $J_\mathrm{Gaussian}(\nu)$ on $\sigma_\mathrm{th}$, the assumed gas temperature has no effect on the emerging Ly$\alpha$ line profile (we use $T=10^4$~K in Fig. \ref{fig_profiles}).
$\sigma_r\simeq(v_r/c)\nu_\alpha$ describes the Doppler broadening of the line due to the rotation of the galaxy. The rotation velocity of the galaxy $v_r$ is closely linked to the halo rotational velocity $v_c= (3\pi G H_0)^{1/3} \Omega_m^{1/6} (1+z)^{1/2} M_h^{1/3}$, ranging between $v_r=v_c$ and $v_r=2v_c$ \citep{mo1998, cole2000}. We assume $v_r=1.5 v_c$.

\subsubsection{Single, double or triple-peaked in a clumpy/homogeneous medium:}
\label{subsubsec_Lya_line_model_clumpy}

This model describes the Ly$\alpha$ line profile emerging from a clumpy medium. It implements the regimes and characteristic escape frequencies identified in  \citet{Gronke2017}. 
We consider a slab with a thickness of $2B$ and a total optical depth of $2\tau_0$. The source is located at the slab's midplane and injects photons at the Ly$\alpha$ resonance $x=0$. If the slab medium is homogeneous, \citet{neufeld1990} derived the emergent Ly$\alpha$ profile as
\begin{equation}
    J_\mathrm{slab}(T,\tau_0, x) = 
        4 \pi \frac{\sqrt{6}}{24} \frac{x^2}{a(T)~ \tau_0} \frac{1}{\cosh\left( \sqrt{\frac{\pi^4}{54}} \frac{|x^3|}{a(T)~\tau_0}\right)},
\label{eq_J_Neufeld}
\end{equation}
for $a(T)\tau_0\gtrsim10^3$ with $a(T)=\frac{A_\alpha}{4\pi \sigma_\mathrm{th}(T)}$ and $\int_{-\infty}^{\infty} J_\mathrm{slab}(x,x_i)\ \mathrm{d}x = 1$. $A_\alpha$ is the Einstein for the spontaneous emission of Ly$\alpha$ photons.

In the following we will revisit the regimes for Ly$\alpha$ escape in a clumpy medium that have been identified in \citet{Gronke2017}. 
The clumpy medium is characterised by the total optical depth of the clumps and the average number of clumps each Ly$\alpha$ photon escaping the slab scatters with. For a slab consisting of clumps with each having an optical depth $\tau_\mathrm{0,cl}$ at the line centre, Ly$\alpha$ photons escaping the slab will encounter on average $f_c$ clumps and have a total optical depth of $\tau_0=\frac{4}{3} f_c \tau_\mathrm{0,cl}$\footnote{The factor $4/3$ arises from the mean path length through a sphere.} at the line centre. The emerging Ly$\alpha$ line profile depends sensitively on the total and clump optical depth at line centre, $\tau_0$ and $\tau_{0,cl}$, respectively, and the number of clumps the Ly$\alpha$ photons scatter with. \citet{Gronke2017} identified the following regimes: 
\begin{itemize}
    \item {\it Free-streaming regime:} The clumpy medium is optically thin ($\tau_\mathrm{0}<1$), and Ly$\alpha$ photons can stream through. The emerging line profile peaks around $x=0$.
    \item {\it Porous regime:} The clumps are optically thick to Ly$\alpha$ photons ($\tau_\mathrm{cl}>1$), but only a fraction $1-\exp(-\tau_{0,cl})$ of the Ly$\alpha$ photons scatter with a clump. The emerging line profile is again peaked around $x=0$.
    \item {\it Random walk regime:} The clumps are optically thick to Ly$\alpha$ ($\tau_\mathrm{cl}>1$), and each Ly$\alpha$ photon encounters $N_\mathrm{sct,rw}\propto f_c^2$ scattering events \citep{Hansen_Oh2006}. However, the number of scattering events is too low for the Ly$\alpha$ photons to scatter in frequency space far enough into the wings to escape through excursion. Hence, the emerging line profile peaks also around $x=0$.
    \item {\it Homogeneous regime:} The clumps are optically thin ($\tau_\mathrm{cl}\leq1$) and Ly$\alpha$ photons scatter $\sim\tau_0$ times ($N_\mathrm{sct,exc}\propto f_c$) and escape via excursion: they follow a random walk in space {\it and} frequency and escape as they are scattered into the wings where the clumps become optically thin. The emerging line profile is a double-peaked with the two peaks being located at 
    \begin{equation}
        x_\mathrm{esc}(\tau_0) \simeq
        \begin{cases}
            \pm \left(k a \tau_0/\sqrt{\pi} \right)^{1/3}, &  \frac{\sqrt{\pi} x_\star^2}{k a} \leq \tau_0 \\
            \pm~ x_\star, & \tau_0 < \frac{\sqrt{\pi} x_\star^2}{k a} 
        \end{cases}
    \end{equation}
    for an injection frequency $x=0$ \citep[c.f.][]{Adams1975, Gronke2017}. 
    Here $x_\star$ is the frequency where the Ly$\alpha$ absorption profile transitions from the Gaussian core to the Lorentzian wings, and $\tau_0 = \frac{\sqrt{\pi} x_\star^2}{k a}$ marks the transition where the slab becomes optically thin at the escape frequency $x_\mathrm{esc}$.
\end{itemize}

\citet{Gronke2017} derived the boundary criteria between these regimes for a static clumpy medium, which we briefly revisit here. To derive the critical number of clumps, $f_c$, separating the regimes, we first consider the time and distances covered that it takes a Ly$\alpha$ to traverse the slab.

{\it Excursion:} 
As Ly$\alpha$ photons traverse or escape the slab, they scatter with \HI many times. This alters their direction and frequency $x$, and they essentially perform a random walk. However, as the Ly$\alpha$ cross section is higher close to the line centre, most scatterings will occur close to the line centre and remain spatially close. Only as the Ly$\alpha$ photons are scattered into the wings of the Ly$\alpha$ absorption profile their mean free paths become larger, allowing them to escape the slab \citep{Adams1975}. The series of these so-called wing scatterings that allow Ly$\alpha$ photons to escape are referred to as excursion.
We can estimate the mean displacement and time spent in such an excursion event: a Ly$\alpha$ photon with frequency $x$ will scatter on average $N_\mathrm{sct,exc}\sim x^2$ times before it returns to the core. For a slab of thickness $B$, its average mean free path is $\lambda_\mathrm{mfp,exc}(x) = B \sigma_0/ (k\tau_0\sigma_\mathrm{HI}(x)) = B / (k \tau_0 H(a,x))$ using the wing approximation of the Ly$\alpha$ cross section and $k\simeq\sqrt{3}$ being a geometrical factor that accounts for slant paths in a plane-parallel medium and was determined in \citet{Adams1975}.\footnote{The geometrical factor $k$ is often not explicitly included in the literature when displacements and excursion times are discussed.} This and the random walk nature of the Ly$\alpha$ escape imply an average displacement of
\begin{eqnarray}
    d_\mathrm{exc} &=& \sqrt{N_\mathrm{sct,exc}} \lambda_\mathrm{mfp,exc}(x) = \frac{\sqrt{N_\mathrm{sct,exc}}B}{k\tau_0 H_v(a,x)} = \frac{x B}{k\tau_0 H_v(a,x)} \nonumber \\
\end{eqnarray}
and time spent in the excursion of
\begin{eqnarray}
    t_\mathrm{exc} &=& N_\mathrm{sct,exc} \frac{\lambda_\mathrm{mfp,exc}(x)}{c} = \frac{N_\mathrm{sct,exc} B}{c k\tau_0 H_v(a,x)} = \frac{x^2 B}{c k\tau_0 H_v(a,x)}, \nonumber \\
\end{eqnarray}
with $H_v(a,x)=\frac{a}{\sqrt{\pi}(x)^2}$ being the effective line absorption profile in the wings. 

{\it Random Walk:} 
As the clumps become optically thick at corresponding escape frequencies $x_\mathrm{esc}(\tau_0)$, the Ly$\alpha$ photons do not escape the slab via excursion anymore but by random walking: the number of scattering events is smaller than required for excursion and scale with the square of the number of clumps, $N_\mathrm{sct,rw}\propto f_c^2$ \citep{Hansen_Oh2006}. With the mean free path given by the average clump separation $\lambda_\mathrm{mfp,rw}=kB/f_c$, the average displacement and time are then
\begin{eqnarray}
    d_\mathrm{rw} &=& \sqrt{N_\mathrm{sct,rw}} \lambda_\mathrm{mfp,rw} = \frac{\sqrt{N_\mathrm{sct,rw}} k B}{f_c} = k B
\end{eqnarray}
and
\begin{eqnarray}
    t_\mathrm{rw} &=& \sqrt{N_\mathrm{sct,rw}} \frac{\lambda_\mathrm{mfp,rw}}{c} = \frac{\sqrt{N_\mathrm{sct,rw}} k B}{c f_c} = \frac{k B f_c}{c}.
\end{eqnarray}

\begin{enumerate}
\item {\it Division between random-walk and homogeneous regime in optically thick medium:}
For a given total optical depth at the line centre, $\tau_0$, we can derive the critical number of clumps along a line of sight that marks the transition from the random (clumps are optically thick) to the homogeneous regime (clumps become optically thick at the excursion frequency). We estimate this transition to arise when both regimes contribute equally to the flux of escaping Ly$\alpha$ photons.
\begin{eqnarray}
    \frac{F_\mathrm{rw}}{F_\mathrm{exc}} &=& \frac{t_\mathrm{exc}}{t_\mathrm{rw}} = \frac{x^2}{k^2 \tau_0 H(x) f_c} = \frac{\sqrt{\pi} x^4}{k^2 a \tau_0 f_c} = 1
\label{eq_FrwDivFexc}
\end{eqnarray}
With $\tau_0=4/3 f_c \tau_\mathrm{0,cl}$, the critical number of clumps for Ly$\alpha$ photons escaping at frequency $x$ yields then as
\begin{eqnarray}
    f_{c,\mathrm{crit}} &=& \frac{\sqrt{3}\pi^{1/4}}{2k} \frac{x^2}{\sqrt{a \tau_\mathrm{0,cl}}}.
\end{eqnarray}
As long as the wings remain optically thick, the majority of Ly$\alpha$ photons (with injection frequency $x=0$) will escape at $x_\mathrm{esc} \simeq \left( \frac{k a \tau_0}{\sqrt{\pi}} \right)^{1/3}$
leading to
\begin{eqnarray}
    f_{c\mathrm{,crit}} &=& 
        \frac{2}{\sqrt{3}k\pi^{1/4}} \sqrt{a \tau_\mathrm{0,cl}}
    \label{eq_fccrit}
\end{eqnarray}
This $f_{c\mathrm{,crit}}$ value marks the transition from the random walk to the excursion regime. We can understand its increase with the clump optical depth $\tau_\mathrm{0,cl}$ as follows: for optically thicker clumps to become optically thin at the escape frequency $x_\mathrm{esc}$, a higher escaping frequency and thus a higher effective total optical depth are required. This can only be achieved by interacting with more clumps (higher $f_{c,\mathrm{crit}}$).

Because the transition described by $f_{c,\mathrm{crit}}$ is not sharp, we model the Ly$\alpha$ line profile emerging from the moving slab by superposing the Ly$\alpha$ radiation escaping in the homogeneous ($J_\mathrm{slab}$) and random walk regimes ($J_\mathrm{centre}$). 
\begin{equation}
    J_\mathrm{rh}(\tau_0, x) = (1 - f_\mathrm{rw}) J_\mathrm{slab}(T, \tau_0, x) + f_\mathrm{rw} J_\mathrm{centre}(T, x)
    \label{eq_Jrh}
\end{equation}
Here we assume $J_\mathrm{centre}$ is given by Eqn. \ref{eq_Jcentre} with $\sigma_r = \sigma_\mathrm{th}$ and 
\begin{equation}
    J_\mathrm{slab}(T,\tau_0, x) = 
    \begin{cases}
        4 \pi \frac{\sqrt{6}}{24} \frac{x^2}{a(T)~ \tau_0} \frac{1}{\cosh\left( \sqrt{\frac{\pi^4}{54}} \frac{|x^3|}{a(T)~\tau_0}\right)}, & \frac{\sqrt{\pi} x_\star^2}{k a(T)} \leq \tau_0\\
        4 \pi \frac{\sqrt{6}}{24} \frac{k x^2}{\sqrt{\pi} x_\star^3} \frac{1}{\cosh\left( \sqrt{\frac{\pi^3}{54}} \frac{k |x^3|}{x_\star^3}\right)}, & \tau_0 < \frac{\sqrt{\pi} x_\star^2}{k a(T)}.
    \end{cases}
\label{eq_Jslab}
\end{equation}
We derive the corresponding ratio $f_\mathrm{fw}$ by assuming that the Ly$\alpha$ flux escapes predominantly where the Ly$\alpha$ profiles peak,
\begin{eqnarray}
    f_\mathrm{rw} &=& \frac{F_\mathrm{rw} / F_\mathrm{exc}}{F_\mathrm{rw} / F_\mathrm{exc} + \frac{J_\mathrm{centre}(0)}{2(J_\mathrm{slab}(\tau_0,x_\mathrm{esc}) + J_\mathrm{slab}(\tau_0,-x_\mathrm{esc}))}}.
\label{eq_frh}
\end{eqnarray}
We note that $J_\mathrm{slab}$ reproduces the results of Ly$\alpha$ radiative transfer simulations down to $\tau_0\simeq10^5$. While the line profiles for lower $\tau_0$ values start deviating, we will see in Section \ref{subsec_optical_depth_HI_column_density} that the galaxies considered here exceed this threshold. Importantly, we find that the assumed $J_\mathrm{slab}$, $J_\mathrm{centre}$ and $f_\mathrm{rw}$ reproduce the Ly$\alpha$ line profiles for resting clumps, fixed $\tau$ values, and varying $f_c$ values in \citet{Gronke2017}.

\item {\it Division between porous and homogeneous regime in optically thin medium:}
As the medium becomes optically thinner, Ly$\alpha$ photons that scatter into the wings can escape the slab before completing their excursion. This transition occurs as the wings become optically thin, i.e. $k\tau(x_\star)\leq1$ translating to $k a \tau_0\leq\sqrt{\pi} x_\star^2$, and Ly$\alpha$ photons escape at $x_\mathrm{esc}=x_\star$. 
While the slab is optically thin at $x_\mathrm{\star}$, depending on whether the clumps are optically thin or thick at $x_\mathrm{esc}$, the escape of Ly$\alpha$ photons is described by the homogeneous and porous regime, respectively. Again we estimate the transition to arise when both regimes contribute equally to the flux of escaping Ly$\alpha$ photons.
\begin{eqnarray}
    \frac{F_\mathrm{por}}{F_\mathrm{hom}} &=& \frac{t_\mathrm{exc}}{t_\mathrm{rw}} = \frac{\sqrt{\pi} x_\star^4}{k^2 a \tau_0 f_c} = 1
\end{eqnarray}
We yield the critical number of clumps that mark the transition from the porous to the homogeneous regime as
\begin{eqnarray}
f_{c\mathrm{,crit}} &=& \frac{x_\star}{k (1-e^{-\tau_\mathrm{0,cl}})}.
\end{eqnarray}
We note that if clumps are optically thin at line centre ($\tau_\mathrm{0,cl}<1$), not every clump encounter leads to a scattering event; this reduces the number of clumps encountered by a factor $1-e^{-\tau_\mathrm{0,cl}}$. 
The emerging Ly$\alpha$ line profile accounts again for Ly$\alpha$ photons escaping in homogeneous ($J_\mathrm{slab}$, see Eqn. \ref{eq_Jslab}) and porous regime ($J_\mathrm{centre}$, see Eqn. \ref{eq_Jcentre}). 
\begin{eqnarray}
    J_\mathrm{ph}(\tau_0, x) &=& (1 - f_\mathrm{por}) J_\mathrm{slab}(T, \tau_0, x) + f_\mathrm{por} J_\mathrm{centre}(T, x)
\label{eq_Jph}
\nonumber \\
\end{eqnarray}
The ratio between the two different escape regimes is then again given by assuming that most Ly$\alpha$ photons escape at the peak frequencies,
\begin{eqnarray}
    f_\mathrm{por} &=& \frac{F_\mathrm{por} / F_\mathrm{hom}}{F_\mathrm{por} / F_\mathrm{hom} + \frac{J_\mathrm{centre}(0)}{2(J_\mathrm{slab}(\tau_0,x_\mathrm{esc}) + J_\mathrm{slab}(\tau_0,-x_\mathrm{esc})}}. \nonumber \\
\label{eq_fph}
\end{eqnarray}

\item {\it Division between porous and random-walk regime:}

As the optical depth of the already optically thick clumps, $\tau_\mathrm{0,cl}$, exceeds the optical depth of the slab, $\tau_0$, a fraction of the Ly$\alpha$ photons traverse the slab without scattering, leaving the random and entering the porous regime. This transition occurs as the number of clumps encountered by the Ly$\alpha$ photons becomes less than unity, $f_c<1$.

\end{enumerate}

\subsubsection{Ionising escape fraction dependent in a clumpy/homogeneous medium}
\label{subsubsec_Lya_line_model_porous}

For this Ly$\alpha$ line profile model, we assume a model similar to the so-called picket fence model \citep{Heckman2011}. Here a fraction $f_\mathrm{esc}$ of the ionising radiation escapes through low-density channels, while the other fraction of ionising photons is absorbed by the dense shell. Correspondingly, the Ly$\alpha$ photons escaping through the channels scatter only a few times, while those escaping through the shell encounter many scattering events. For optically thin channels, the former gives rise to a single-peaked Ly$\alpha$ line centred around $x=0$, while the latter creates a broader double-peaked Ly$\alpha$ line (assuming a homogeneous slab model with peaks at $x_\mathrm{esc}$). 

Here we assume the channels to be fully ionised and a fraction $f_\mathrm{esc}$ of the Ly$\alpha$ photons to escape through them without scattering. As a consequence, the other fraction of Ly$\alpha$ photons encounter the shell with an optical depth of 
\begin{eqnarray}
    \tau_\mathrm{shell} &=& \frac{\tau_0}{1 - f_\mathrm{esc}},
    \label{eq_tau_shell}
\end{eqnarray}
at their first scattering. $\tau_0$ is the optical depth as derived in Section \ref{subsec_optical_depth_HI_column_density}. We make the simplifying assumption that those photons traverse the shell without being scattered into the channels. In reality a fraction of these photons encounter a lower optical depth when traversing the empty channels on their scattering path out of the slab, leading to a profile closer centered around $x=0$. Thus, the emerging Ly$\alpha$ line profile, which we assume to be
\begin{eqnarray}
    J(\tau, x) &=& f_\mathrm{esc}\  J_\mathrm{slab}(T, 0, x) \\
    && +\ \left(1 - f_\mathrm{esc}\right)\  J_\mathrm{shell}(T, \tau_\mathrm{shell}, x) \nonumber \\
    J_\mathrm{shell} &=& f_\mathrm{shell} J_\mathrm{slab}(T, 0, x) \\
    && +\ \left(1 - f_\mathrm{shell}\right)\  J_\mathrm{slab}(T, \tau_\mathrm{shell}, x) \nonumber
\end{eqnarray}
represents a lower limit to the fraction of Ly$\alpha$ photons escaping close to the resonance. We derive $f_\mathrm{shell}$ by choosing a clump optical depth $\tau_\mathrm{0,cl}$ and use Eqn \ref{eq_frh} and \ref{eq_fph} for the random-homogeneous ($ka\tau_0 > \sqrt{\pi} x_\star^3$) or porous-homogeneous ($ka\tau_0 < \sqrt{\pi} x_\star^3$) transitions.

\subsection{Optical depth and \HI column density}
\label{subsec_optical_depth_HI_column_density}

To derive the Ly$\alpha$ line profile emerging from a simulated galaxy for the models described in Sections \ref{subsubsec_Lya_line_model_clumpy} ({\it Clumpy} model) and \ref{subsubsec_Lya_line_model_porous} ({\it Porous} model), we yield the optical depth at the Ly$\alpha$ line centre from our simulated galaxies as
\begin{eqnarray}
    \tau_0 &=& \frac{4}{3} f_c \tau_\mathrm{0,cl} = N_\mathrm{HI} \sigma_\mathrm{HI}.
\end{eqnarray}
$\tau_\mathrm{0,cl}$ is a free parameter in our model and reflects the optical depth of a dense clump in the ISM. 
We will use this parameter to calibrate our model to the observed Ly$\alpha$ LFs at $z=6.6-8$ in Section \ref{sec_number_and_properties_LAEs}. To obtain a rough estimate of $\tau_\mathrm{0,cl}$, we consider the median mass ($M_\mathrm{cl}$) and size ($r_\mathrm{cl}$) of molecular clouds ($M_\mathrm{cl}\simeq 10^5\msun$ and $r_\mathrm{cl}=20$~pc) resembling the mass and size of possible dense structures in the ISM,
\begin{eqnarray}
    \tau_\mathrm{0,cl} &=& \sigma_\mathrm{HI} \frac{M_\mathrm{cl}}{r_\mathrm{cl}}.
\end{eqnarray}
To obtain the optical depth $\tau_0$, we derive the neutral hydrogen column density $N_\mathrm{HI}$ from the initial gas mass, $M_g^\mathrm{i}$ as
\begin{eqnarray}
    N_\mathrm{HI} &=& \xi \frac{3 M_\mathrm{HI}}{4\pi r_g^2 m_\mathrm{H}} = \xi \frac{3 X_c (1-Y) M_g^\mathrm{i}}{4\pi\ (4.5\lambda r_\mathrm{vir})^2 m_\mathrm{H}} = \xi \frac{3 f_m M_\mathrm{vir}}{81\pi \lambda^2 r_\mathrm{vir}^2 m_\mathrm{H}} \nonumber \\
    &=& \xi \left( \frac{9\pi^2 H_0^2 \Omega_m}{G} \right)^{2/3} (1+z)^2 M_\mathrm{vir}^{1/3} \frac{3 f_m}{81\pi \lambda^2 m_\mathrm{H}}.
\end{eqnarray}
Here $r_g$ describes the gas radius, for which we assume $r_g= 4.5 \lambda r_\mathrm{rvir}$. $X_c$ and $Y$ are the cold gas and helium mass fractions, respectively. Gas accretion and SN feedback processes determine the relation between the initial gas mass and the halo mass, $f_m$, which ranges typically between $\sim10^{-3}$ for low-mass galaxies to $\sim 10^{-1}$ for more massive galaxies. 
$\xi$ is a geometrical correction factor that depends on $\tau_0$ and the dust optical depth at the Ly$\alpha$ resonance $\tau_d$. Its maximum values is $0.35$ and we describe its derivation and dependencies in Appendix \ref{app_correction_factor}.
For the cosmological parameters assumed in this paper, we yield
\begin{eqnarray}
    N_\mathrm{HI} = 6.5\times 10^{17} \mathrm{cm}^2\ (1+z)^2 \frac{\xi f_m}{\lambda^2} \left( \frac{M_\mathrm{vir}}{10^8 M_\odot} \right)^{1/3}.
\end{eqnarray}

\subsection{Dust attenuation}
\label{subsec_dust_attenuation}

We employ two different dust models. The first one links the Ly$\alpha$ escape fraction to the escape fraction of UV continuum photons, $f_\mathrm{esc}^\mathrm{c}$.
The second one is more complex. It assumes a clumpy medium where the attenuation of Ly$\alpha$ by dust follows different relations in the regimes identified in \citet{Gronke2017}. Both models assume a slab-like geometry and we describe their details in the following.

\subsubsection{Simple attenuation model}
\label{subsubsec_dust_simple_model}

In this model, we assume that (i) dust and gas are perfectly mixed, (ii) the dust distribution is slab-like, and (iii) the dust attenuation of Ly$\alpha$ photons is proportional to the dust attenuation of UV continuum photons. The escape fraction of Ly$\alpha$ photons, $f_\mathrm{esc}^\mathrm{Ly\alpha}$, is then directly related to the escape of UV continuum photons, $f_\mathrm{esc}^\mathrm{c}$, derived in Section \ref{subsubsec_metals_and_dust}.
\begin{eqnarray}
    f_\mathrm{esc}^\mathrm{Ly\alpha} &=& p\ f_\mathrm{esc}^\mathrm{c}
\end{eqnarray}
We use $p$ as a free parameter to obtain the observed Ly$\alpha$ luminosity functions at $z=6.6-7.3$.

\subsubsection{Refined attenuation model}
\label{subsubsec_dust_refined_model}

This model assumes that dust and gas are perfectly mixed and distributed in clumps. The dust attenuation of Ly$\alpha$ photons depends on the total optical depth of the dust, $\tau_\mathrm{d,total}$, the optical depth of a clump, $\tau_\mathrm{d,cl}$, and the number of clumps, $f_c$, encountered along the sightline from the midplane to the surface of the slab.
We derive its value by estimating the dust absorption cross section. Following \citet{Galliano2022} and assuming the radius and density of graphite/carbonaceous grains (see Section \ref{subsubsec_metals_and_dust}), we assume $\kappa_\mathrm{abs} \simeq \frac{Q_\mathrm{abs}}{a s} \simeq 2\times10^5$~cm$^{2}/$g with $Q_\mathrm{abs}\simeq1$ being the absorption efficiency. \footnote{We note that this is in rough agreement with the dust extinction cross sections of the Small and Large Magellanic clouds $\kappa_\mathrm{ext} = \sigma_\mathrm{d} / m_\mathrm{H} = \sigma_\mathrm{d,ref} \frac{M_z}{M_\mathrm{d}\ Z_\mathrm{ref} m_\mathrm{H}}  \simeq 4\times 10^5$~cm$^{2}/$g, with the extinction efficiency $Q_\mathrm{ext}=Q_\mathrm{abs}+Q_\mathrm{sca}$ being given by the similar sized absorption ($Q_\mathrm{abs}$) and scattering efficiencies ($Q_\mathrm{sca}$) at Ly$\alpha$, a dust-to-metal mass ratio $M_\mathrm{d}/M_Z\simeq0.25$, $\sigma_\mathrm{ref}\simeq4\times10^{-22}$cm$^2$ and $Z_\mathrm{ref}\simeq0.0025$ for SMC and $\sigma_\mathrm{ref}\simeq7\times10^{-22}$cm$^2$ and $Z_\mathrm{ref}\simeq0.005$ for LMC \citep[for further explanations see][]{Laursen2010}.}
\begin{eqnarray}
    \tau_\mathrm{d,total} &=& \frac{4}{3} f_c \tau_\mathrm{d,cl} 
    = \xi\ \frac{3}{4\pi} \frac{M_\mathrm{d}}{r_\mathrm{d}^2} \kappa
    = \frac{M_\mathrm{d}}{M_\mathrm{HI}} \frac{\kappa_\mathrm{abs} m_\mathrm{H}}{\sigma_\mathrm{HI}} \tau_0
\end{eqnarray}
The resulting estimates for $\tau_\mathrm{d,total}$ and $\tau_\mathrm{d,cl}$ allow us to compute the Ly$\alpha$ escape fractions in the different escape regimes as follows. 

\paragraph{Free-streaming regime:}
\label{par_dust_freestreaming}

In an optically thin slab ($\tau_0<1$), the Ly$\alpha$ photons stream through $\sim f_c$ clumps. On their way, they are attenuated by the dust in clumps and hence, the total dust optical depth determines the Ly$\alpha$ escape fraction, $\tau_\mathrm{d,total}$, as
\begin{eqnarray}
    f_\mathrm{esc}^\mathrm{Ly\alpha, fs} &=& \exp\left(-\tau_\mathrm{d, total}\right) = \exp\left(-\frac{4}{3} f_c \tau_\mathrm{d,cl}\right)
\end{eqnarray}
We note that in this regime, the number of clumps along the sightline $f_c$ and clump optical depth $\tau_\mathrm{0,cl}$ are degenerate.

\paragraph{Random walk regime:}
\label{par_dust_randomwalk}

In the random walk regime, both the slab and individual clumps are optically thick ($\tau_\mathrm{cl}\geq1$). As a result, Ly$\alpha$ photons escape by mostly being scattered by the clumps, and their escape fraction is determined by the number of clumps encountered along their random walk, $N_\mathrm{cl}(f_c)$, and the absorption probability per clump interaction $\epsilon$. According to \citet{Hansen_Oh2006}, it is then given by
\begin{eqnarray}
    f_\mathrm{esc}^\mathrm{Ly\alpha, rw} &=& f_\mathrm{HO06} = \frac{1}{\cosh(\sqrt{2 N_\mathrm{cl}(f_c)\ \epsilon})}
\end{eqnarray}
We assume $N_\mathrm{cl}(f_c)\simeq \frac{3}{2}f_c^2 + 2 f_c$ as found in \citet{Gronke2017}. The scaling of $N_c$ with $f_c$ also agrees with the findings in \citet{Hansen_Oh2006} and prefactors vary slightly due to different geometries of the scattering surface. However, since $\epsilon$ is sensitive to how deep the photons permeate the clump, it depends non-trivially on the clump optical depth and movement. For simplicity, we assume $\epsilon=1-\exp(-\tau_\mathrm{d,cl}^{2/3} / 10)$, which we have found to be in rough agreement with the fits and results shown in \citet{Gronke2017}.

\paragraph{Homogeneous regime:}
\label{par_dust_homogeneous}

In the homogeneous regime, the slab is optically thick ($\tau_0\geq1$), while the individual clumps are optically thin at the escape frequencies ($\tau_\mathrm{cl}(x_\mathrm{esc})<0$). During their initial random walk, the Ly$\alpha$ photons scatter with $N_\mathrm{cl}(f_{c,\mathrm{crit}})$ clumps before they diffuse into the wings and escape by free-stream through $f_c$ clumps.
The resulting Ly$\alpha$ escape fraction,
\begin{eqnarray}
    f_\mathrm{esc}^\mathrm{Ly\alpha, hom} &=& f_\mathrm{HO06}(f_{c,\mathrm{crit}}) \exp(-\tau_\mathrm{d, total}),
\end{eqnarray}
depends on $f_{c,\mathrm{crit}}$ and $\tau_\mathrm{d,total}$, with $f_{c,\mathrm{crit}}$ being determined by $\tau_0$ and $\tau_\mathrm{0,cl}$. 
\begin{eqnarray}
f_{c,\mathrm{crit}} =
\begin{cases}
    \frac{2}{\sqrt{3}k \pi^{1/4}} \sqrt{a \tau_\mathrm{0,cl}} & \mathrm{for}\ ka\tau_0 \geq \sqrt{\pi} x_\mathrm{max}^2\\
    \frac{x_\mathrm{max}}{k \left(1-e^{-\tau_\mathrm{0,cl}}\right)} & \mathrm{for}\ ka\tau_0 < \sqrt{\pi} x_\mathrm{max}^2
\end{cases}
\end{eqnarray}

\paragraph{Porous regime:}
\label{par_dust_porous}

In the porous regime, the individual clumps are optically thick ($\tau_\mathrm{0,cl}\geq1$), but only a fraction $1-e^{-f_c}$ of the Ly$\alpha$ photons will encounter a clump along their sightlines. The other fraction of Ly$\alpha$ photons does not interact with any clumps and is thus not attenuated by dust as they escape the slab.\footnote{We note that our expression is here a lower limit of $f_\mathrm{esc}^\mathrm{Ly\alpha, por}$ as we assume the Ly$\alpha$ radiation interacting with clumps to experience attenuation as if they streamed through the clump. It might be more appropriate to consider these Ly$\alpha$ photons to be absorbed as in the random walk regime, $f_\mathrm{esc}^\mathrm{Ly\alpha, por} = e^{-f_c} + \left[1 - e^{-f_c}\right]  \frac{1}{\cosh(\sqrt{2 N_\mathrm{cl}(f_c)\ \epsilon})}$, however in practise galaxies in the porous regime have not much, if any, dust.}
\begin{eqnarray}
    f_\mathrm{esc}^\mathrm{Ly\alpha, por} &=& e^{-f_c} + \left[1 - e^{-f_c}\right] \exp\left(-\frac{4}{3} f_c \tau_\mathrm{d,cl}\right)
\end{eqnarray}

\subsubsection{Emerging Ly$\alpha$ line profile models}
\label{subsubsec_emerging_Lya_profile_models}

We briefly summarize the combinations of Ly$\alpha$ line and dust attenuation models that we will investigate in this paper.

\paragraph*{Gaussian:} 

The Ly$\alpha$ line profile emerging from a galaxy is given by the central Gaussian Ly$\alpha$ line profile (Section \ref{subsubsec_Lya_line_model_gaussian}). To account for the attenuation by dust, we apply the Ly$\alpha$ escape fraction, $f_\mathrm{esc}^\mathrm{Ly\alpha}$, derived in our simple dust model (Section \ref{subsubsec_dust_simple_model}) to all frequencies $x$. 

\paragraph*{Clumpy:} 

This model assumes an shell of dusty gas clumps, whereas gas and dust are perfectly mixed. It combines the Ly$\alpha$ line model described in Section \ref{subsubsec_Lya_line_model_clumpy} with the refined dust model depicted in Section \ref{subsubsec_dust_refined_model}. The gas in the galaxies is assumed to have a temperature of $T=10^4$~K.\footnote{We have chosen $T=10^4$~K for simplicity. If we were to assume the virial temperature ($T_\mathrm{vir}$), the double-peak line profile would narrow as $T_\mathrm{vir}$ increases.} In contrast to the {\it Gaussian} model, we dust-attenuate the Ly$\alpha$ line of each escape regime (homogeneous, random, porous) by its corresponding escape fraction $f_\mathrm{esc}^\mathrm{Ly\alpha}$. The emerging Ly$\alpha$ line  profile is then the superposition of the line profiles of all relevant escape regimes,
\begin{equation}
    L_\alpha^\mathrm{gal}(x) = f_\mathrm{esc, slab}^\mathrm{Ly\alpha} (1-f) J_\mathrm{slab}(x) + f_\mathrm{esc, centre}^\mathrm{Ly\alpha} f J_\mathrm{centre}(x),
\end{equation}
with $f_\mathrm{esc, slab}^\mathrm{Ly\alpha}=f_\mathrm{esc, hom}^\mathrm{Ly\alpha}$, and $(f, f_\mathrm{esc, centre}^\mathrm{Ly\alpha})$ given by $(1, f_\mathrm{esc, fs}^\mathrm{Ly\alpha})$, $(f_\mathrm{rw}, f_\mathrm{esc, rw}^\mathrm{Ly\alpha})$ or $(f_\mathrm{por}, f_\mathrm{esc, por}^\mathrm{Ly\alpha})$ depending on the total and clump optical depths $\tau_0$ and $\tau_\mathrm{0,cl}$.

\paragraph*{Porous:} 

This model is very similar to the {\it Clumpy} model. However, it considers the shell of clumps to be pierced with gas and dust-free channels through which a fraction $f_\mathrm{esc}$ of the Ly$\alpha$ photons escape without scattering. It combines the Ly$\alpha$ line model described in Section \ref{subsubsec_Lya_line_model_porous} and assuming $\tau_\mathrm{channel}^\mathrm{LyC}=0$ with the refined dust model depicted in Section \ref{subsubsec_dust_refined_model}. Again we assume the gas in the galaxy to be heated to a temperature of $T=10^4$~K, and the Ly$\alpha$ line of each escape regime (homogeneous, random, porous) to be dust-attenuated by its corresponding escape fraction $f_\mathrm{esc}^\mathrm{Ly\alpha}$. The emerging Ly$\alpha$ line profile is again a superposition of the Ly$\alpha$ photons escaping through the channels and the clumpy shell,
\begin{eqnarray}
    L_\alpha^\mathrm{gal}(x) &=& f_\mathrm{esc}~ J_\mathrm{channel}(x)\ +\ (1- f_\mathrm{esc})~ J_\mathrm{shell}(x) \\
    J_\mathrm{channel}(x) &=& J_\mathrm{centre}(x)\\
    J_\mathrm{shell}(x) &=& f_\mathrm{esc, slab}^\mathrm{Ly\alpha}~ (1-f)~ J_\mathrm{slab}(x) \ +\ f_\mathrm{esc, centre}^\mathrm{Ly\alpha}~ f~ J_\mathrm{centre}(x) \nonumber \\
\end{eqnarray}
with $f_\mathrm{esc, slab}^\mathrm{Ly\alpha}=f_\mathrm{esc, hom}^\mathrm{Ly\alpha}$, and $(f, f_\mathrm{esc, centre}^\mathrm{Ly\alpha})$ given by $(1, f_\mathrm{esc, fs}^\mathrm{Ly\alpha})$, $(f_\mathrm{rw}, f_\mathrm{esc, rw}^\mathrm{Ly\alpha})$ or $(f_\mathrm{por}, f_\mathrm{esc, por}^\mathrm{Ly\alpha})$ depending on the total and clump optical depths $\tau_0$ and $\tau_\mathrm{0,cl}$. We note that $\tau_0$ exceeds the $\tau_0$ value in the {\it Clumpy} model when $f_\mathrm{esc}>0$ (see Eqn. \ref{eq_tau_shell}), as the same amount of gas and dust is distributed over a smaller solid angle.

\subsection{IGM attenuation}
\label{subsec_IGM_attenuation}

The Ly$\alpha$ radiation escaping from a galaxy is attenuated by the \HI  it encounters along the line of sight from the location of emission, $r(z_\mathrm{em})$, to the location of absorption, $r(z_\mathrm{obs})$. Expressing the frequency $\nu$ of a photon in terms of its rest-frame velocity $x = v/b = (\nu_\alpha/\nu - 1) c/b$ relative to the Ly$\alpha$ line centre, the transmitted fraction of radiation at frequency $x$ is given by
\begin{eqnarray}
T_{\alpha,x}(x) &=& \exp \left[ -\tau_\alpha(x) \right] \\
\tau_\alpha(x) &=& c \int_{z_\mathrm{em}}^{z_\mathrm{obs}} \sigma_0~ \phi(x + x_\mathrm{p}(r(z)))\ \frac{n_\mathrm{HI}(r(z))}{(1+z) H(z)}\ \mathrm{d}z.\
\label{eq_IGMtransmission_tau_alpha}
\end{eqnarray}
Here $\tau_\alpha$ describes the optical depth to Ly$\alpha$, while $n_\mathrm{HI}(r)$ and $v_\mathrm{p}(r)= b x_\mathrm{p}(r)$ the \HI density and peculiar velocity (in the rest-frame of the emitted Ly$\alpha$ radiation) at a physical distance $r$ from the emitter, respectively. $\sigma_0$ is the specific absorption cross section, described in the cgs system as
\begin{eqnarray}
\sigma_0 &=& \frac{\pi e^2 f}{m_e c^2} = \frac{3 \lambda_\alpha^2 A_{21}}{8\pi},
\end{eqnarray}
where $f=0.4162$ is the oscillator strength, $e$ the electron charge, $m_e$ the electron mass, $\lambda_\alpha=1216$\AA~ the wavelength of a photon at the Ly$\alpha$ line centre, and $A_{21}=6.265\times10^8$s$^{-1}$ the Einstein coefficient for spontaneous emission of Ly$\alpha$ photons.
$\phi(x)$ depicts the Ly$\alpha$ profile for absorption and is given by a Voigt profile consisting of a Gaussian core
\begin{eqnarray}
\phi_\mathrm{Gauss}(x) &=& \frac{\lambda_\alpha}{\sqrt{\pi} b} ~\exp\left( - x^2 \right) \\
b &=&  \sqrt{\frac{2 k_B T_\mathrm{IGM}}{m_H}},
\end{eqnarray}
and Lorentzian damping wings
\begin{eqnarray}
\phi_\mathrm{Lorentz}(x) &=& \frac{A_{21} \lambda_\alpha^2}{4\pi^2 (x~b)^2 + \frac{1}{4} A_{21}^2 \lambda_\alpha^2}.
\end{eqnarray}
Here $b$ is the Doppler parameter, $T_\mathrm{IGM}$ the temperature of the IGM, $k_B$ the Boltzmann constant, and $m_\mathrm{H}$ the mass of a hydrogen atom. While pressure line broadening is unimportant in regions of low \HI density and the profile can be approximated by the Gaussian core, the absorption in the Lorentzian damping wings is non-negligible in regions of high \HI density. In practise, we mimic the Voigt profile by assuming the Gaussian core profile $\phi(x)=\phi_\mathrm{Gauss}(x)$ for $|x|<x_\star$ and the Lorentzian wing profile $\phi(x)=\phi_\mathrm{Lorentz}(x)$ otherwise. Fitting to numerical results yields the transition frequency as
\begin{eqnarray}
x_\star &=& 0.54 \log_{10}\left(\frac{b}{\mathrm{cm/s}}\right)
\end{eqnarray}
for temperatures between $T=0.01$K and $10^8$K. 

Our calculations of $T_\alpha$ include the Hubble flow and peculiar velocities $v_\mathrm{p}$: outflows (inflows) of gas from a galaxy that correspond to positive (negative) $v_\mathrm{p}$ values will redshift (blueshift) the Ly$\alpha$ photons and lead to an increase (decrease) in $T_\alpha$.
For each galaxy in a simulation snapshot we derive $T_\alpha$ along all directions along the major axes (i.e. along and against the x, y and z axes). By stepping through the simulation box that is divided into $512$ cells on the side (and each cell having a size of $461$ckpc), we derive the $n_\mathrm{HI}(r)$ and $v_\mathrm{p}(r)$ profiles from the {\sc astraeus} ionisation and {\sc vsmdpl} density and velocity grids. For any galaxy, we start the profiles at the galaxy position $r_\mathrm{em}=0$ and end them once the highest frequency $x_\mathrm{max} = v_\mathrm{max}/b=40$ tracked in our Ly$\alpha$ line profiles has redshifted out of absorption at $r\simeq v_\mathrm{max}/ [H_0 \Omega_m^{1/2}(1+z)^{1/2}] \simeq 13.6/(1+z)^{1/2}$cMpc. We assume $T_\mathrm{IGM}=10^4$K in ionised and $T_\mathrm{IGM}=10^2$K in neutral regions.
Since the Ly$\alpha$ line redshifts out of resonance very quickly (the light travel time for distance $r$ at $z=7$ is less than $2$~Myrs, shorter than the simulation time steps), a single simulation snapshot suffices for computing the $T_\alpha$ values of the galaxies in that snapshot. We also assume periodic boundary conditions when computing $T_\alpha$.

Finally, we derive the observed, i.e. dust and IGM attenuated, Ly$\alpha$ luminosity and line profile along each major axes (resulting in 6 lines of sight) as
\begin{eqnarray}
    L_{\alpha,x}(x) &=& L_\alpha^\mathrm{gal}(x)\ T_{\alpha,x}(x) = L_\alpha^\mathrm{intr}\ f_\mathrm{esc}^\mathrm{Ly\alpha}\ J(x),
\end{eqnarray}
where $f_\mathrm{esc}^\mathrm{Ly\alpha}$ and $J(x)$ are the respective Ly$\alpha$ escape fraction and line profile for a one of the models as outlined in Section \ref{subsubsec_emerging_Lya_profile_models}. The total observed Ly$\alpha$ luminosity $L_\alpha$ and total fraction of Ly$\alpha$ radiation transmitted through the IGM are yielded when integrating the respective quantity over the frequency $x$.
\begin{eqnarray}
    L_\alpha &=& \int_{-\infty}^{\infty} L_{\alpha,x}(x)\ \mathrm{d}x \\
    T_\alpha &=& \int_{-\infty}^{\infty} T_{\alpha,x}(x)\ \mathrm{d}x
\end{eqnarray}
In the following, we use all lines of sight as independent probes when line-of-sight-sensitive Ly$\alpha$ quantities are analysed.

We derive the observed Ly$\alpha$ luminosities ($L_\alpha$) for all galaxies at $z=20$, $15$, $12$, $10$, $9$, $8$, $7.3$, $7$ and $6.6$ for any combination of emerging Ly$\alpha$ line model ({\it Gaussian}, {\it Clumpy}, {\it Porous}) and reionisation scenario ({\sc mhdec}, {\sc mhconst}, {\sc mhinc}). Free model parameters ($p$ for the {\it Gaussian} model, $\tau_\mathrm{0,cl}$ for the {\it Clumpy} and {\it Porous} models) have been chosen to visually best-fit the observed Ly$\alpha$ LFs at $z\simeq6.7$, $7.0$ and $7.3$ (see Tab. \ref{tab_Lya_line_and_ionisation_topology_models}). For simplicity and better comparison we assume in all models the gas in galaxies to have the temperature of photo-ionised gas, $T=10^4$~K.
Moreover, we note that since the {\sc mhconst} scenario represents an intermediate case and provides no further insights, we limit our discussion to the {\sc mhdec} and {\sc mhinc} scenarios in the remainder of this paper.

\begin{table}
    \centering
    \begin{tabular}{c|c|c|c|c}
         Parameter & Scenario & {\it Gaussian} & {\it Clumpy} & {\it Porous} \\
         \hline
         $\tau_\mathrm{0,cl}$ & {\sc mhdec} & - & $1.2\times10^6$ & $2.4\times10^6$ \\
         $\tau_\mathrm{0,cl}$ & {\sc mhinc} & - & $5\times10^5$ & $1.8\times10^6$ \\
         $p$ & {\sc mhdec} & 1.0 & - & - \\
         $p$ & {\sc mhinc} & 1.4 & - & - \\
         $T$ & all & $10^4$~K & $10^4$~K & $10^4$~K \\
    \end{tabular}
    \caption{Parameters for our three different Ly$\alpha$ line profile models}
    \label{tab_Lya_line_and_ionisation_topology_models}
\end{table}

\section{Numbers and properties of Ly$\alpha$ emitting galaxies}
\label{sec_number_and_properties_LAEs}

In this Section, we aim to identify which physical process -- the intrinsic Ly$\alpha$ production ($L_\alpha^\mathrm{intr}$), the absorption by dust within the galaxies ($f_\mathrm{esc}^\mathrm{Ly\alpha}$), or the scattering by \HI in the IGM ($T_\alpha$) -- dominates the observed Ly$\alpha$ emission. To this end, we analyse (i) how the IGM attenuation profile $T_\alpha(x)$ depends on galaxy mass and the $f_\mathrm{esc}$-sensitive ionisation topology, (ii) how the Ly$\alpha$ line profiles emerging from a galaxy depend on the density and velocity distributions of gas and dust within a galaxy and $f_\mathrm{esc}$, and how much it affects the fraction of Ly$\alpha$ radiation that is transmitted through the IGM, and (iii) to which degree the $f_\mathrm{esc}$ dependency of $L_\alpha^\mathrm{intr}$, $f_\mathrm{esc}^\mathrm{Ly\alpha}$, and $T_\alpha$ leave characteristic imprints in the Ly$\alpha$ luminosity functions and the population emitting visible Ly$\alpha$ emission.

\begin{figure*}
    \centering
    \includegraphics[width=0.99\textwidth]{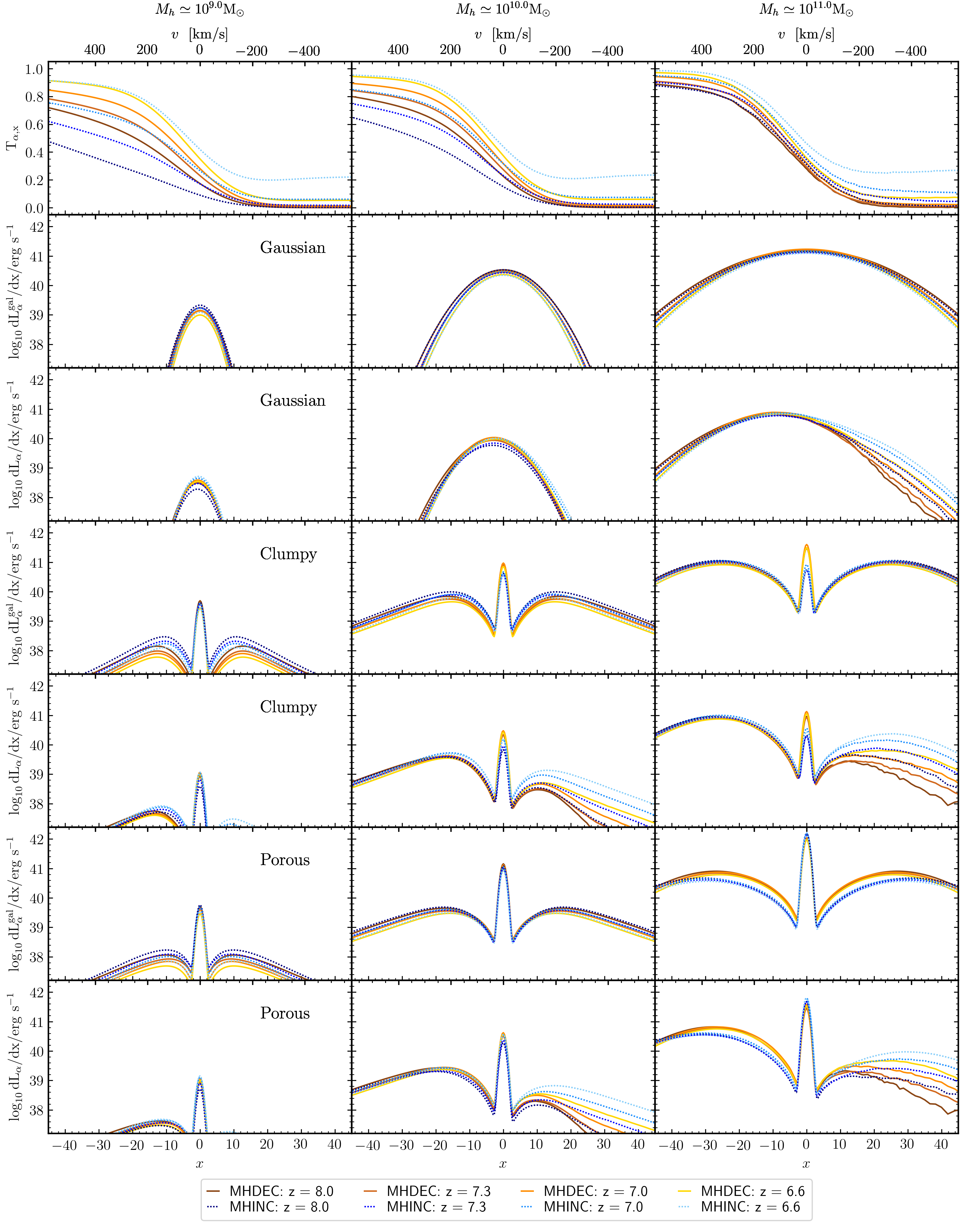}
    \caption{Intrinsic (top) and observed (bottom) Ly$\alpha$ line profile and IGM transmission (centre) at $z=8.0$, $7.3$, $7.0$, $6.6$ for a homogeneous static gas shell (left), a clumpy gas shell (centre), and a clumpy gas shell with holes through which Ly$\alpha$ radiation escapes without scattering. Solid (dashed dotted) lines show results for the reionisation scenario where $f_\mathrm{esc}$ decreases (increases) with halo mass $M_h$.}
    \label{fig_profiles}
\end{figure*}

\subsection{The transmission through the IGM}
\label{subsec_discussion_IGM_transmission}

We start by discussing the frequency-dependent IGM transmission $T_{\alpha,x}$ shown in the top row of Fig. \ref{fig_profiles}. These profiles depend solely on the underlying ionisation topology and density distribution of the IGM. From the different panels depicting the average $T_{\alpha,x}$ in different halo mass bins of width $\Delta\log_{10}M_h=0.125$, we see that all $T_{\alpha,x}$ profiles follow a common trend: $T_{\alpha,x}$ decreases towards higher frequencies with an stronger decline around the Ly$\alpha$ resonance ($x=0$). Photons bluewards the Ly$\alpha$ resonance redshift into the Ly$\alpha$ resonance as they propagate through the IGM and have the largest likelihood to be absorbed by the \HI present. Photons redwards the Ly$\alpha$ resonance are also redshifted, but their likelihood of being absorbed by \HI decreases significantly as their energy drops.

In each panel in the top row of Fig. \ref{fig_profiles} we show $T_{\alpha,x}$ for the two reionisation scenarios {\sc mhdec} (yellow/orange/brown lines) and {\sc mhinc} (blue lines) and redshifts $z=8.0$, $7.3$, $7.0$, $6.6$ (bright to dark lines as redshift decreases). In general, i.e. for both reionisation scenarios and all halo masses, $T_{\alpha,x}$ increases as the ionised regions grow around galaxies and the IGM is increasingly ionised (bright to dark lines): firstly, a larger ionised region shifts not only the point of strongest Ly$\alpha$ absorption to higher frequencies $x$ but also reduces the absorption in the damping wings of the Ly$\alpha$ absorption profile. Secondly, lower \HI fractions in ionised regions diminish the number of \HI atoms absorbing Ly$\alpha$ photons.

These two mechanisms shape $T_{\alpha,x}$ redwards and bluewards the Ly$\alpha$ resonance. As Ly$\alpha$ photons travel through the IGM and redshift, photons emitted at frequencies $x\gtrsim0$ see the Gaussian core of the Ly$\alpha$ absorption profile $\phi(x)$ and are absorbed by \HI abundances as low as $\chi_\mathrm{HI}\gtrsim10^{-4}$; thus they are sensitive to the residual \HI fraction in ionised regions. Correspondingly, we see in Fig. \ref{fig_profiles} that $T_{\alpha,x}$ increases for $x\gtrsim0$ with decreasing redshift as the photoionisation rate around galaxies increases and lowers the residual \HI fraction in ionised regions.
However, photons emitted at frequencies $x\lesssim 0$ are absorbed by the damping wings of the Ly$\alpha$ absorption profile $\phi(x)$. Since the Ly$\alpha$ absorption cross section is lower in the damping wings, the abundance of \HI needs to be significantly higher for Ly$\alpha$ photons to be absorbed; thus, as the sizes of ionised regions decrease, photons emitted at these frequencies are increasingly absorbed by the neutral regions located beyond the ionised regions around the emission sites. For this reason, we find $T_{\alpha,x}$ for $x\lesssim0$ to increase as the sizes of the ionised regions around galaxies rise with increasing halo mass and decreasing redshift. The rising sizes of ionised regions also become manifest in the shift of the frequency at which $T_{\alpha,x}$ has a value of $0.5$ to higher frequencies.

Its dependence on the size of the ionised regions around galaxies makes $T_{\alpha,x}$ a tracer of the ionisation topology: our two extreme reionisation scenarios where $f_\mathrm{esc}$ either increases ({\sc mhinc}, blue dotted lines) or decreases ({\sc mhdec}, yellow to brown solid lines) with rising halo mass $M_h$ exhibit very different ionisation topologies (see Fig. \ref{fig_XHImaps_with_LAEs}). These differences are imprinted in $T_{\alpha,x}$ as follows. 
Firstly, since in the {\sc mhinc} scenario the higher $f_\mathrm{esc}$ values of more massive galaxies ($M_h\gtrsim10^{10}\msun$) raise the photoionisation rate within ionised regions (leading to lower $\chi_\mathrm{HI}$ values, also seen in Fig. \ref{fig_hist_ion} at $z\lesssim6$), the corresponding $T_{\alpha,x}$ values are higher bluewards the Ly$\alpha$ resonance than in the {\sc mhdec} scenario. Moreover, in the {\sc mhinc} scenario, reionisation proceeds faster, leading to the Universe being more ionised at $z<7$, and the bias of the ionising emissivity towards more massive galaxies grows with time, raising the photoionisation rate in the ionised regions. Both effects contribute to the relative increase in $T_{\alpha,x}$ from {\sc mhdec} to {\sc mhinc} to rise towards lower redshifts bluewards the Ly$\alpha$ resonance.
Secondly, as the size of the ionised regions around galaxies is imprinted in $T_{\alpha,x}$ redwards the Ly$\alpha$ resonance, {\sc mhinc} shows lower (higher) $T_{\alpha,x}$ values at $z\gtrsim7$ ($z\lesssim7$) than the {\sc mhdec} scenario for galaxies with $M_h<10^{11}\msun$: 
At $z\gtrsim7$, ionised regions become increasingly smaller towards lower mass halos ($M_h\lesssim10^{9.5}\msun$) and higher redshifts as the corresponding $f_\mathrm{esc}$ values and global ionisation fraction decrease. However, at $z\lesssim7$, this trend reverses as the ionised regions become large enough for the red wing of the Ly$\alpha$ to be redshifted out of the absorption resonance of the Gaussian core. Towards more massive halos and higher global ionisation fractions, $T_{\alpha,x}$ becomes sensitive to the residual \HI fraction in ionised regions (c.f. $T_{\alpha,x}$ in {\sc mhinc} (light blue dotted line) exceeds $T_{\alpha,x}$ in {\sc mhdec} (yellow solid line) at $z=6.6$). It is interesting to note that the respective $T_{\alpha,x}$ values are very similar in both reionisation scenarios, despite the $f_\mathrm{esc}$ values of more massive halos ($M_h>10^{10}\msun$) differing by about one order of magnitude or more.

\begin{figure*}
    \centering
    \includegraphics[width=0.99\textwidth]{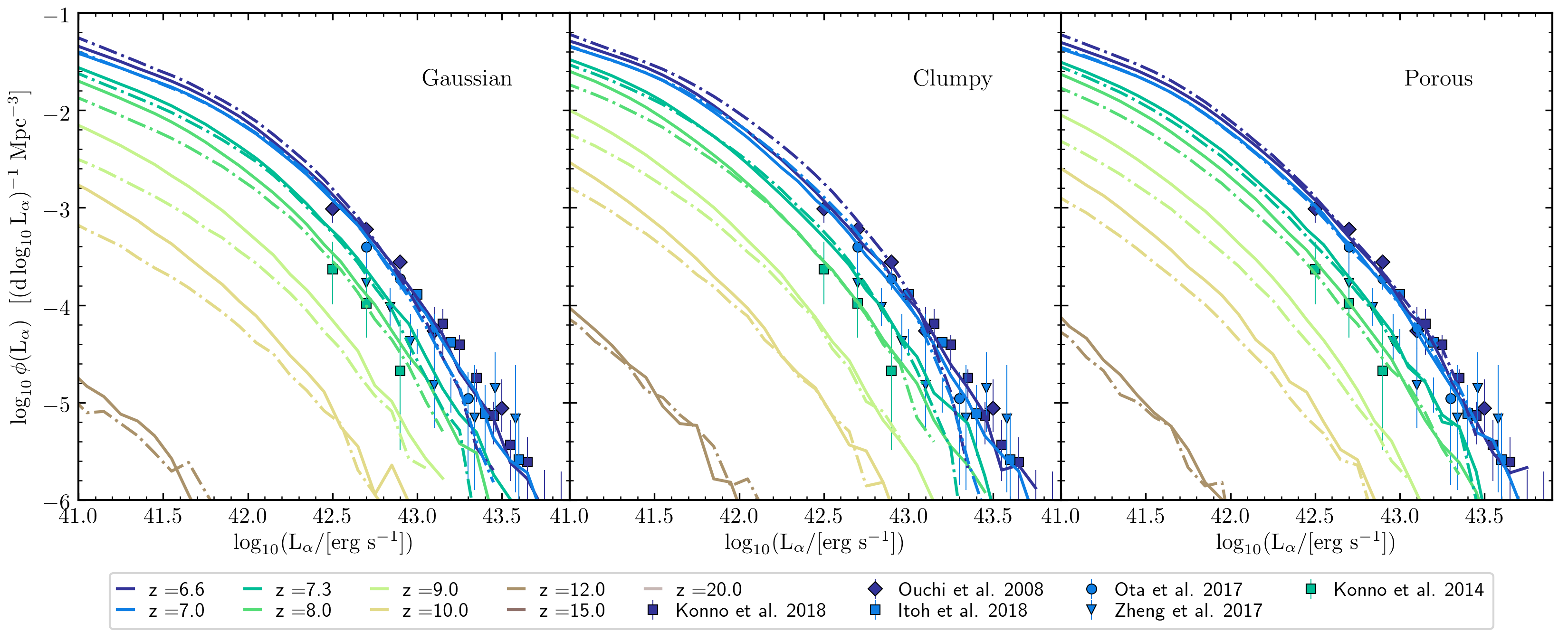}
    \caption{Observed Ly$\alpha$ luminosity functions at $z=20$, $15$, $12$, $10$, $9$, $8$, $7.3$, $7$, $6.6$ for a homogeneous static gas shell (left), a clumpy gas shell (centre), and a clumpy gas shell with holes through which Ly$\alpha$ radiation escapes without scattering. Solid (dashed dotted) lines show results for the reionisation scenario where $f_\mathrm{esc}$ decreases (increases) with halo mass $M_h$. Observational data points are from \citet{Ouchi2010, Konno2014, Ota2017, Zheng2017, Konno2018, Itoh2018}.}
    \label{fig_LyaLF}
\end{figure*}

\subsection{The Ly$\alpha$ line profiles and luminosity functions}

The Ly$\alpha$ line profile emerging from a galaxy represents a quantity that (i) is shaped by the density and velocity distribution of gas and dust within the galaxy and (ii) affects which fraction of the Ly$\alpha$ radiation escaping from a galaxy is transmitted through the IGM. In this Section, for our three models of the emerging Ly$\alpha$ line profiles, we discuss the following: (i) How do the assumed gas and dust distributions affect the attenuation of Ly$\alpha$ by dust in a galaxy and the emerging Ly$\alpha$ line profile? (ii) How does the Ly$\alpha$ line profile affect the Ly$\alpha$ transmission through the IGM? And, since the luminosity function of the intrinsic Ly$\alpha$ luminosity ($L_\alpha^\mathrm{intr}$) will be steeper for the scenario where $f_\mathrm{esc}$ increases ({\sc mhinc}) than when it decreases ({\sc mhdec}) with rising halo mass, (iii) which characteristics are required for the Ly$\alpha$ line profiles of the simulated galaxy population to reproduce the observed Ly$\alpha$ luminosity functions (Ly$\alpha$ LFs)?

\subsubsection{The Gaussian model}

The {\it Gaussian} line model centers the Ly$\alpha$ line at the Ly$\alpha$ resonance. The second row in Fig. \ref{fig_profiles} shows that its width increases as the rotational velocity of a galaxy increases with rising halo mass. Both the increase in the line width and the size of the ionised region surrounding the galaxy lead to higher IGM transmission values of Ly$\alpha$ radiation as galaxies become more massive (c.f. third row in Fig. \ref{fig_profiles}). At the same time, the fraction of Ly$\alpha$ photons that escape from the galaxies drops as the abundance of dust increases. We use the ratio between the Ly$\alpha$ and UV continuum escape fractions to adjust the Ly$\alpha$ luminosities emerging from the galaxies and fit the observed Ly$\alpha$ LFs in each of our reionisation scenarios. In the {\sc mhinc} scenario the more massive galaxies -- that dominate the observed Ly$\alpha$ LF -- have higher $f_\mathrm{esc}$ values than in the {\sc mhdec} scenario; to compensate the corresponding lower $L_\alpha^\mathrm{intr}$ values (and steeper slope of the intrinsic Ly$\alpha$ LF), we need a higher $f_\mathrm{esc}^\mathrm{Ly\alpha}/f_\mathrm{esc}^\mathrm{UV}$ ratio ($1.4$) than in the {\sc mhdec} scenario ($1.0$). Despite this compensation, the slopes of the observed Ly$\alpha$ LFs at $z\lesssim8$ (c.f. left panel in Fig. \ref{fig_LyaLF}) is still steeper for the {\sc mhinc} than for the {\sc mhdec} scenario.

\subsubsection{The Clumpy model} 

In the {\it Clumpy} model, the clumpiness of the gas in the shell and the attenuation by dust molecules in these clumps determine the shape of the Ly$\alpha$ line profile. We note that in the following clumpiness describes the number of clumps in the dusty gas shell, i.e. a higher clumpiness corresponds to fewer clumps and thus a higher ratio between the clump ($\tau_\mathrm{0,cl}$) and total line optical depth ($\tau_0$).
We find the following characteristic trends for the Ly$\alpha$ line profile: 
Firstly, the clumpier the gas in the shell is, the more Ly$\alpha$ radiation escapes around the Ly$\alpha$ resonance (profile showing a central peak), and the fewer Ly$\alpha$ photons escape through excursion or via the wings (double peak profile). 
Secondly, when assuming the same clump size for all galaxies -- as we do in this paper -- the gas clumpiness decreases as galaxies become more massive and contain more gas. Thus, from low-mass to more massive galaxies, we find the Ly$\alpha$ line profile to shift from a central peak dominated to a double-peak domintaed profile (see the fourth row in Fig. \ref{fig_profiles} from left to right), reflecting the transition from the random to the homogeneous regime (see Section \ref{subsubsec_Lya_line_model_clumpy}). This transition also goes in hand with an increased transmission through the IGM, which we can see when comparing the Ly$\alpha$ profiles emerging from galaxies (fourth row) with those after having traversed the IGM (fifth row in Fig. \ref{fig_profiles}). The Ly$\alpha$ luminosity at $x=0$ decreases by $\sim0.5$ orders of magnitude for all halo masses (from $10^{41.6}$erg~s$^{-1}$ to $10^{41.1}$erg~s$^{-1}$ for $M_h\simeq10^{11}\msun$ and from $10^{39.7}$erg~s$^{-1}$ to $10^{39.2}$erg~s$^{-1}$ for $M_h\simeq10^{9}\msun$ for e.g. {\sc mhdec} model), while the peak Ly$\alpha$ luminosity of the red wing decreases only about $\lesssim0.3$ orders of magnitude at all halo masses. While the blue wing is similarly or more attenuated than the central peak in the IGM, the total fraction of Ly$\alpha$ radiation transmitted through the IGM for a fully-double peaked profile exceeds that of profiles with a central peak component.
Furthermore, as the galaxies' gravitational potentials flatten with decreasing redshift, $\tau_0$ decreases and leads to (i) a narrower double-peak profile (following the dependence of the peak position on $\tau_0^{1/3}$) and (ii) a stronger central peak (the gas becomes clumpier as the ratio $\tau_\mathrm{0,cl}/\tau_0$ increases).

A change in the clumpiness of the gas and dust shell (or clump optical depth $\tau_\mathrm{0,cl}$ and $\tau_\mathrm{d,cl}$) goes not only in hand with a change in the Ly$\alpha$ profile affecting $T_\alpha$ but also an altered attenuation of the escaping Ly$\alpha$ radiation by dust. Thus, adjusting the clump optical depth allows us to enhance and reduce the Ly$\alpha$ luminosities and reproduce the observed Ly$\alpha$ LFs: As we increase the size of the clumps, i.e. increase $\tau_\mathrm{0,cl}$, Ly$\alpha$ photons will scatter with fewer clumps, leading to (i) a higher fraction $f_\mathrm{esc}^\mathrm{Ly\alpha}$ escaping, and (ii) a higher fraction escaping at the Ly$\alpha$ resonance, which again leads to stronger attenuation by \HI in the IGM.
However, we note that once the emerging Ly$\alpha$ profile is fully double-peaked, the attenuation by \HI in the IGM can not be further decreased (by changing the injected Ly$\alpha$ line profile). The observable Ly$\alpha$ emission can only be enhanced by decreasing $\tau_\mathrm{0,cl}$ as long as the $f_\mathrm{OH06}(f_\mathrm{c,crit})$ factor in $f_\mathrm{esc}^\mathrm{Ly\alpha,hom}$ remains significantly below unity. With the observed Ly$\alpha$ LF being dominated by the more massive galaxies ($M_h\gtrsim10^{10}\msun$, as we will discuss in the next Section), we find the $\tau_\mathrm{0,cl}$ value to reflect the factor by which the bright end of the intrinsic Ly$\alpha$ LF needs to be reduced to reproduce the observational Ly$\alpha$ LF data points (filled points in Fig. \ref{fig_LyaLF}).
As the intrinsic Ly$\alpha$ LFs is lower at the bright end in the {\sc mhinc} scenario, a lower $\tau_\mathrm{0,cl}$ value ($5\times10^5$) is required than for the {\sc mhdec} scenario ($1.2\times10^6$). Nevertheless, the slopes of the resulting observed Ly$\alpha$ LFs at $z\lesssim8$ keep the trends of the intrinsic Ly$\alpha$ LFs, with the bright ends of the Ly$\alpha$ LFs being steeper in the {\sc mhinc} than in the {\sc mhdec} scenario.

\subsubsection{The Porous model} 

The {\it Porous} model represents a refinement of the {\it Clumpy} model. It adds gas-free channels through which Ly$\alpha$ and ionising photons escape freely. This explains why, to first order, we find the trends in the last two rows of Fig. \ref{fig_profiles} to be similar to those in the fourth and fifth rows: a lower clumpiness of gas and dust in the shell induces a stronger prevalence of the double-peak component in the Ly$\alpha$ line profile emerging from a galaxy, enhancing the IGM transmission $T_\alpha$ and absorption by dust within the galaxy, and causing the corresponding Ly$\alpha$ LFs to shift to lower values.
On the other hand, it differs from the {\it Clumpy} model substantially, as $f_\mathrm{esc}$ determines the minimum fraction of Ly$\alpha$ radiation that escapes at the Ly$\alpha$ resonance and contributes to the central peak in our modelling. Hence, as long as $\tau_\mathrm{0,cl}$ remains above the $\tau_\mathrm{0,cl}$ value that leads to the same fraction of Ly$\alpha$ escaping in the central peak than given by $f_\mathrm{esc}$ (referred to as $\tau_\mathrm{0,cl}^\mathrm{crit}$ in the following), the {\it Porous} model inherits the trend of the {\it Clumpy} model. 
As $\tau_\mathrm{0,cl}$ drops below $\tau_\mathrm{0,cl}^\mathrm{crit}$, a further decrease in $\tau_\mathrm{0,cl}$ affects the Ly$\alpha$ line profile emerging from a galaxy hardly, and once the $f_\mathrm{HO06}(f_\mathrm{c,crit})$ factor of $f_\mathrm{esc}^\mathrm{Ly\alpha,hom}$ approaches unity, the observed Ly$\alpha$ LFs remain "fixed". 
The resulting upper limit of $f_\mathrm{esc}^\mathrm{Ly\alpha}$ (determined by the total dust optical depth $\tau_\mathrm{d,total}$) is essential, as together with $L_\alpha^\mathrm{intr}$ it provides an upper limit to $f_\mathrm{esc}$ values that fit the observed Ly$\alpha$ LFs. We find this upper limit to be about $f_\mathrm{esc}\sim0.5$ in our {\sc astraeus} model.

Due to their opposing dependencies of $f_\mathrm{esc}$ with halo mass, the Ly$\alpha$ profiles in the {\it Porous} model show the most noticeable differences between the {\sc mhdec} and {\sc mhinc} scenarios among our three Ly$\alpha$ line profile models. While the double-peak component is more prominent in the most massive galaxies ($M_h\simeq10^{11}\msun$) in the {\sc mhdec} scenario, the central peak is slightly stronger in the {\sc mhinc} scenario. 
To fit the observed Ly$\alpha$ LFs, we find that we require for both reionisation scenarios a more clumpy gas and dust distribution than in the {\it Clumpy} model, i.e. a (higher) $\tau_\mathrm{0,cl}$ value of $1.8-2.4\times10^6$. These increased $\tau_\mathrm{0,cl}$ values enhance the corresponding $f_\mathrm{c,crit}$ values and thus the dust attenuation in the homogeneous regime giving rise to the double-peak components and counteract the increased escape close to the Ly$\alpha$ resonance.
This model-integrated correlation between $f_\mathrm{esc}^\mathrm{Ly\alpha}$ and $f_\mathrm{esc}$ counteracts the trend of flattening (steepening) the slope of the intrinsic Ly$\alpha$ LFs due to $f_\mathrm{esc}$ decreasing (increasing) with rising halo mass: If $f_\mathrm{esc}$ is low (high), more (less) Ly$\alpha$ radiation is subject to dust attenuation. This model feature explains why the observed Ly$\alpha$ LFs of the {\sc mhinc} simulation are shallower than in the {\it Clumpy} model and hardly changes for the {\sc mhdec} simulation due to its low $f_\mathrm{esc}$ values for more massive galaxies.

\begin{figure*}
    \centering
    \includegraphics[width=0.99\textwidth]{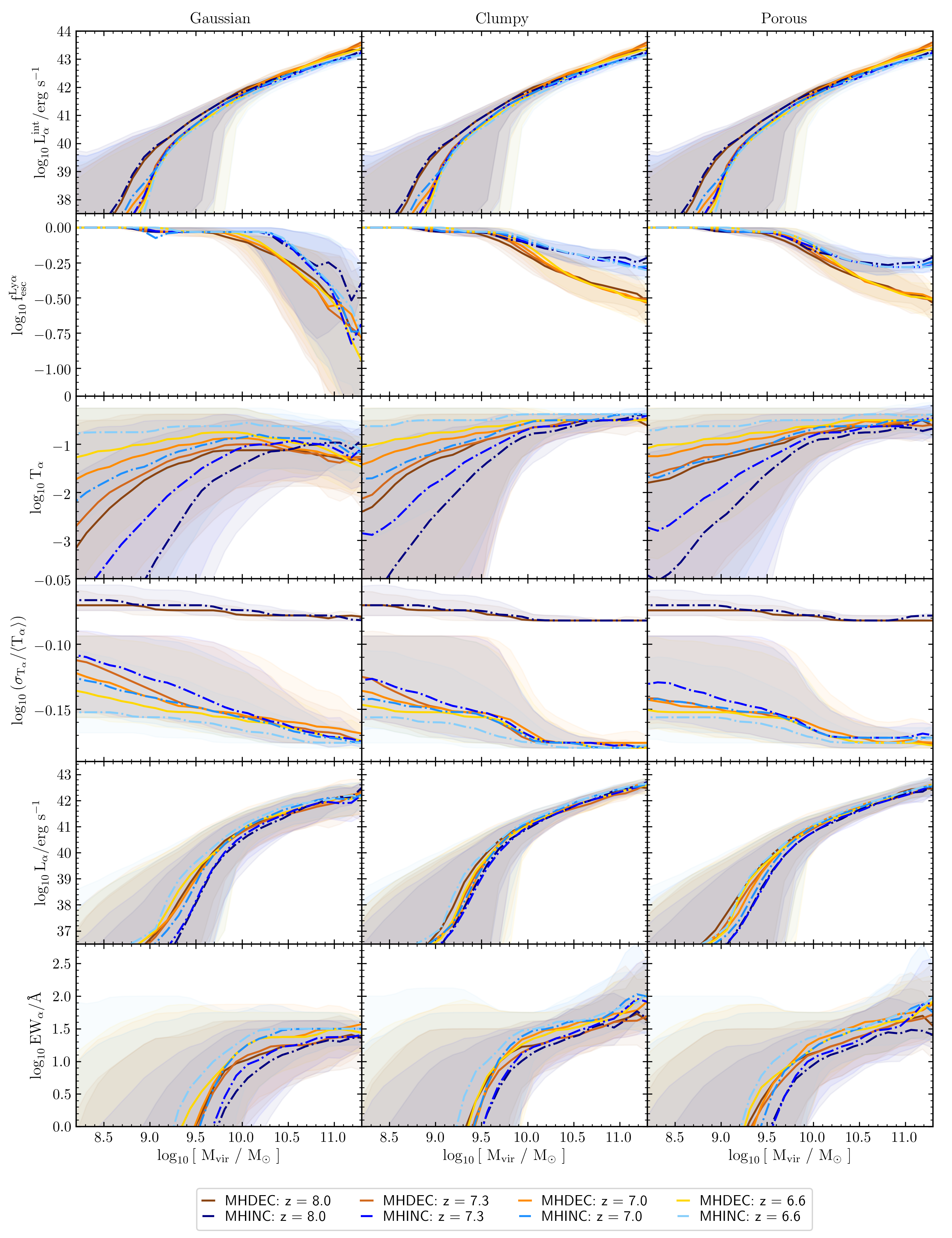}
    \caption{Median of indicated galactic properties (lines) and their $\sim1.3\sigma$ uncertainties (shaded regions) as a function of halo mass $M_h$ at $z=8.0$, $7.3$, $7.0$, $6.6$ for a homogeneous static gas shell. Solid (dashed dotted) lines show results for the reionisation scenario where $f_\mathrm{esc}$ decreases (increases) with halo mass $M_h$.}
    \label{fig_histograms}
\end{figure*}

As the dust composition and absorption cross section for Ly$\alpha$ remain highly uncertain during the EoR, we note that a lower (higher) dust absorption cross section $\kappa_\mathrm{abs}$ could still reproduce the observed Ly$\alpha$ LFs in our {\it Clumpy} and {\it Porous} models by raising (decreasing) the clump optical depth $\tau_\mathrm{0,cl}$. However, this would go along with an enhanced (reduced) double-peak and reduced (enhanced) central-peak component in the average Ly$\alpha$ line profile emerging from galaxies.

Finally, we briefly comment on how our emerging and IGM-attenuated Ly$\alpha$ profiles compare to those obtained from radiative hydrodynamical simulations of clouds and small cosmological volumes ($\sim10^3$cMpc$^3$). While the {\it Clumpy} and {\it Porous} reproduce the double- and triple-peak profiles and their dependence on $N_\mathrm{HI}$ and $f_\mathrm{esc}$ found in cloud simulations \citep{Kakiichi2021, Kimm2019, Kimm2022} by construction, their Ly$\alpha$ line profiles differ from those obtained from the {\sc sphinx} simulation \citep{Garel2021}. 
In {\sc sphinx} the median angle-averaged Ly$\alpha$ line profile has been found to be less double-peaked towards brighter galaxies, with the blue peak being seemingly increasingly suppressed. This is the opposite trend of our findings. The discrepancy lies in the differently assumed or simulated ISM structures: While our LAE models assume an idealised scenario of same-sized dusty gas clumps, the {\sc sphinx} simulation follows the formation of star-forming clouds within galaxies. With rising galaxy mass, we expect the simulated {\sc sphinx} galaxies to contain a higher number of star-forming clouds with various velocity and size distributions. A single or very few star-forming clouds -- as found in low-mass galaxies -- will give rise to a double-peaked Ly$\alpha$ line profile. Adding the profiles of multiple/many star-forming clouds at different velocities will give rise to increasingly more complex Ly$\alpha$ line profiles as galaxies become more massive. Adjusting our current Ly$\alpha$ line profile models to the complex structure of the ISM will be the subject of future work.

\subsection{The dependence of Ly$\alpha$ properties on halo mass}

In this Section, we provide a more detailed discussion of how the intrinsic Ly$\alpha$ luminosity ($L_\alpha^\mathrm{intr}$), the Ly$\alpha$ escape fraction, the Ly$\alpha$ transmission through the IGM, the observed Ly$\alpha$ luminosity, and Ly$\alpha$ equivalent width depend on halo mass and evolve with redshift for the different reionisation scenarios. To this end, we show these quantities as a function of halo mass for both reionisation scenarios ({\sc mhdec}: yellow/orange/brown lines; {\sc mhinc}: blue lines) and redshifts $z\simeq8$, $7.3$, $7$, $6.6$ in Fig. \ref{fig_histograms} and list the corresponding average \HI fractions in Table \ref{table_XHI_reionisation_scenarios}. Solid and dot-dashed lines in Fig. \ref{fig_histograms} depict the median value for galaxies in the given halo mass bin, and shaded regions indicate the range spanned by $68\%$ of the values. For line-of-sight-dependent Ly$\alpha$ properties ($T_\alpha$, $L_\alpha$, EW$_\alpha$), we include all $6$ lines of sight.

\begin{table}
\centering
\begin{tabular}{|c|c|c|}
    $z$ & $\langle\chi_\mathrm{HI}\rangle^\mathrm{MHINC}$ & $\langle\chi_\mathrm{HI}\rangle^\mathrm{MHDEC}$\\
    \hline
     8.0 & 0.84 & 0.71 \\
     7.3 & 0.69 & 0.59 \\
     7.0 & 0.52 & 0.49 \\
     6.6 & 0.23 & 0.34 \\
\end{tabular}
\caption{The evolution of the global \HI fractions of the IGM for our reionisation scenarios.}
\label{table_XHI_reionisation_scenarios}
\end{table}

\paragraph*{Intrinsic Ly$\alpha$ luminosity $L_\alpha^\mathrm{intr}$:} 

As the most recent star formation dominates the production of ionising photons within galaxies, we find $L_\alpha^\mathrm{intr}$ to follow the SFR-$M_h$ relation \citep[for a detailed discussion, see][]{Hutter2021a}. While the range of SFR values is broad for low-mass halos ($M_h\lesssim10^{9.5}\msun$) where SN feedback drives stochastic star formation, the SFR-$M_h$ relation becomes tighter towards more massive galaxies as SN feedback ejects an increasingly lower fraction of gas from the galaxy. Being mainly produced by recombining hydrogen atoms within a galaxy, the Ly$\alpha$ radiation produced within the galaxy correlates with the escape fraction of ionising photons as $1-f_\mathrm{esc}$. As we can see from the first row in Fig. \ref{fig_histograms}, this dependency on $f_\mathrm{esc}$ leads to higher (lower) Ly$\alpha$ luminosities for more massive galaxies, lower (higher) Ly$\alpha$ luminosities for low-mass galaxies, and thus a shallower (steeper) LFs in the {\sc mhdec} ({\sc mhinc}) scenario.

\paragraph*{Ly$\alpha$ escape fraction $f_\mathrm{esc}^\mathrm{Ly\alpha}$:}

As the dust content in galaxies increases with their mass, we find $f_\mathrm{esc}^\mathrm{Ly\alpha}$ to decrease with rising halo mass at all redshifts and for all Ly$\alpha$ line models. However, the different assumed distributions of dust and their resulting attenuation of Ly$\alpha$ radiation lead to differences in the details of this global trend: 
Firstly, the {\it Gaussian} model shows a steeper decline in $f_\mathrm{esc}^\mathrm{Ly\alpha}$ for galaxies with $M_h\gtrsim10^{10.5}\msun$ than the {\it Clumpy} and {\it Porous} models.
Secondly, $f_\mathrm{esc}^\mathrm{Ly\alpha}$ is always higher in the {\sc mhinc} than in the {\sc mhdec} scenario. This is necessary to reproduce the observed Ly$\alpha$ LFs by compensating the lower intrinsic Ly$\alpha$ luminosities with a more clumpy gas-dust distribution in the {\sc mhinc} scenario. In case of the {\it Clumpy} model, it also highlights how a decrease in the clump optical depth by a factor $\sim 2$ can increase $f_\mathrm{esc}^\mathrm{Ly\alpha}$ by reducing the fraction of Ly$\alpha$ photons escaping in the homogeneous regime (i.e. a decrease in $f_\mathrm{c,crit}$ and $\tau_\mathrm{0,cl}$ leads to a reduced number of clumps encounters $N_\mathrm{cl}$ and the clump albedo $\epsilon$).
Thirdly, for the {\sc mhdec} ({\sc mhinc}) scenario, the $f_\mathrm{esc}^\mathrm{Ly\alpha}$ values show higher (lower) values in the {\it Porous} model than in the {\it Clumpy} model for $M_h\lesssim10^{10}\msun$. The reason for this difference is as follows. In both scenarios the higher $\tau_\mathrm{0,cl}$ values in the {\it Porous} model increase the dust attenuation of Ly$\alpha$ escaping in the homogeneous regime. But only a fraction $1-f_\mathrm{esc}$ of the Ly$\alpha$ photons is subject to dust attenuation. This unattenuated escape of Ly$\alpha$ radiation imprints the mass-dependency of $f_\mathrm{esc}$ in the $f_\mathrm{esc}^\mathrm{Ly\alpha}$ values. However, for galaxies with $M_h\gtrsim10^{10}\msun$, this imprint ($f_\mathrm{esc}^\mathrm{Ly\alpha}$ enhancement in {\it Porous} model) is only visible in the {\sc mhinc} scenario where $f_\mathrm{esc}$ values are sufficiently large ($>0.1$); in the {\sc mhdec} scenario $f_\mathrm{esc}$ values are too small.

\paragraph*{Ly$\alpha$ IGM transmission $T_\alpha$:}

As outlined in Section \ref{subsec_discussion_IGM_transmission}, the surrounding ionised region (in particular its size and residual \HI fraction) and the  Ly$\alpha$ line profile emerging from a galaxy determine how much of a galaxy's escaping Ly$\alpha$ radiation is transmitted through the IGM.

For more massive galaxies with $M_h\gtrsim10^{10}\msun$, $T_\alpha$ is mainly shaped by the Ly$\alpha$ profile. This is because the ionised regions surrounding them are sufficiently large -- due to their enhanced ionising emissivity and their clustered neighbourhood -- for the Ly$\alpha$ radiation to redshift out of absorption. Hence, at these high halo masses, any trends in $T_\alpha$ reflect the ratio between the Ly$\alpha$ radiation escaping around the Ly$\alpha$ resonance and escaping through the wings: the more Ly$\alpha$ escapes in the central peak, the lower is the $T_\alpha$ value. Indeed, as can be seen in Fig. \ref{fig_histograms}, the {\it Gaussian} model concentrating the emerging Ly$\alpha$ radiation around the Ly$\alpha$ resonance shows the lowest median $T_\alpha$ values at $M_h\gtrsim10^{10}\msun$ among all Ly$\alpha$ line profile models.
In the {\it Clumpy} model, where the fraction of Ly$\alpha$ escaping through the wings increases with rising halo mass, we find the median $T_\alpha$ value to increase accordingly. This effect is more evident for the {\sc mhinc} scenario as it transitions from a Ly$\alpha$ line profile with a dominating central peak at $M_h\simeq10^{10}\msun$ to one with a prevailing double peak component at $M_h\simeq10^{11}\msun$.
The {\it Porous} model also confirms that $T_\alpha$ is highly sensitive to the Ly$\alpha$ line profile. In the {\sc mhinc} scenario, the double peak component is weaker and increases less with halo mass, leading to slightly lower $T_\alpha$ values than in the {\it Clumpy} model for $M_h\simeq10^{10-11}\msun$ and $T_\alpha$ hardly changing with halo mass. In the {\sc mhdec} scenario, we see the same effect but to a lower degree.

However, for less massive galaxies ($M_h\lesssim10^{10}\msun$), $T_\alpha$ is more sensitive to the properties of their surrounding ionised regions. Since the ionised regions around less massive galaxies can differ significantly depending on their environment and phase in their stochastic star formation cycle (see \citet{Hutter2021b} and \citet{Legrand2023} for environment dependence), their $T_\alpha$ values span across an extensive range from as low as effectively zero to as high as $\simeq70\%$. Nevertheless, the median $T_\alpha$ value shows a definite trend. It increases with rising halos mass for all models and at all stages of reionisation. With increasing halo mass, galaxies are surrounded by larger ionised regions as they form more stars emitting ionising photons and are more likely to be located in clustered regions that are reionised earlier. The larger the surrounding ionised regions are, the higher the transmission of Ly$\alpha$ radiation through the IGM.
We can see this relationship when comparing the median $T_\alpha$ values of the {\sc mhdec} and {\sc mhinc} simulations. In the {\sc mhdec} scenario low-mass galaxies are surrounded by larger ionised regions at $z\gtrsim7$ than in the {\sc mhinc}, causing their corresponding $T_\alpha$ values to be raised (c.f. orange/brown solid lines vs dark blue/blue lines in the third row of Fig. \ref{fig_histograms}). At $z\lesssim7$, however, reionisation progresses faster and the photoionisation rate in clustered ionised regions yields a lower residual \HI fraction in the {\sc mhinc} simulation, both leading to a higher median $T_\alpha$ value for the {\sc mhinc} than {\sc mhdec} scenario at $z\simeq6.6$.
Finally, we briefly discuss how the Ly$\alpha$ line profile emerging from a galaxy affects $T_\alpha$ for less massive galaxies. From Fig. \ref{fig_histograms} we see that the $T_\alpha$ values differ between our three different Ly$\alpha$ line profile models: 
While at all stages of reionisation the $T_\alpha$ values for $M_h\lesssim10^{10}\msun$ are very similar in the {\it Porous} and {\it Clumpy} model, the {\it Porous} model shows lower $T_\alpha$ values for $M_h\gtrsim10^{10}\msun$ at $z\lesssim7$ than the {\it Clumpy} model in the {\sc mhinc} scenario. This drop goes in hand with the increased central peak component in these more massive galaxies (c.f. Fig. \ref{fig_profiles} and the previous Section). 
The median $T_\alpha$ values of the {\it Gaussian} model always lie below those of the {\it Clumpy} and {\it Porous} models; a larger fraction of Ly$\alpha$ radiation escapes closer to the Ly$\alpha$ resonance and is thus subject to stronger attenuation by the IGM.

\paragraph*{Variance of the IGM transmission along different lines of sight:}

To investigate how strongly the transmission of Ly$\alpha$ radiation through the IGM depends on the direction, we show the standard deviation of $T_\alpha$ values over the $6$ lines of sight aligning with the major axes in relation to the corresponding mean value, $\sigma_{T_\alpha} / \langle T_\alpha \rangle = \sqrt{\langle T_\alpha^2\rangle - \langle T_\alpha\rangle^2}/ \langle T_\alpha \rangle$, in the fourth row of Fig. \ref{fig_histograms}.
At all redshifts and for all models, $\sigma_{T_\alpha} / \langle T_\alpha \rangle$ decreases with rising halo mass and decreasing redshift for the following reason. As galaxies grow in mass, they produce more ionising photons that can ionise larger regions around them and are also more likely to be located in more strongly clustered ionised regions, both enhancing and homogenising Ly$\alpha$ transmission through the IGM along different lines of sight. However, we note that parts of the decrease of $\sigma_{T_\alpha} / \langle T_\alpha \rangle$ with decreasing redshift is also due to $\langle T_\alpha \rangle$ rising. Since it is hard to disentangle these two effects, we will focus on relative differences between the different reionisation scenarios and Ly$\alpha$ line profile models. 
Firstly, the more the emerging Ly$\alpha$ line profile is concentrated around the Ly$\alpha$ resonance, the more sensitive is $T_\alpha$ to the varying \HI abundance around a galaxy, and the larger is the variance across different lines of sight (c.f. the higher $\sigma_{T_\alpha} / \langle T_\alpha \rangle$ values in the {\it Gaussian} compared to the other two models, and in the {\it Porous} compared to the {\it Clumpy} model for $M_h\gtrsim10^{10.5}\msun$ when central peak component dominates). 
Secondly, we focus on Ly$\alpha$ line profiles more sensitive to the environmental \HI abundance of a galaxy ({\it Gaussian} model). When accounting for the $\langle T_\alpha \rangle$ values to be lower in the {\sc mhinc} than in the {\sc mhdec} scenario at $z\simeq7$ (see median $T_\alpha$ values in the third row of Fig. \ref{fig_histograms}), we can deduce that the variance of $T_\alpha$ across different lines of sight is higher in the {\sc mhdec} than in the {\sc mhinc} scenario. Indeed in the {\sc mhinc} scenario, the shape of ionised regions is closer to spheres and less filamentary, which results in more ``homogeneous" $T_\alpha$ values.

\paragraph*{Observed Ly$\alpha$ luminosity $L_\alpha$:}

For any model and reionisation scenario, the trend of $L_\alpha$ with rising halo mass depends on the respective trends of $L_\alpha^\mathrm{intr}$, $f_\mathrm{esc}^\mathrm{Ly\alpha}$, and $T_\alpha$. 
Being surrounded by smaller ionised regions, the low $T_\alpha$ values of less massive galaxies ($M_h\lesssim10^{10}\msun$) strongly suppress and shape their emerging Ly$\alpha$ radiation.
In contrast, the $T_\alpha$ values of more massive galaxies ($M_h\gtrsim10^{10}\msun$) show only weak trends with halo mass and similar values throughout reionisation. For this reason, the trends of their $L_\alpha$ values with halo mass are predominantly shaped by the corresponding trends of $L_\alpha^\mathrm{intr}$ and $f_\mathrm{esc}^\mathrm{Ly\alpha}$. Though, for model parameters that reproduce the observed Ly$\alpha$ LFs, a relative increase (decrease) of $L_\alpha^\mathrm{intr}$ towards higher halo masses, such as in the {\sc mhinc} ({\sc mhdec}) scenario, is compensated by an $f_\mathrm{esc}^\mathrm{Ly\alpha}$ that decreases more (less) strongly with halo mass. Nevertheless, the resulting relation between $L_\alpha$ and halo mass does not significantly change. It shows that only more massive galaxies where SN and radiative feedback do not considerably suppress star formation exhibit observable Ly$\alpha$ emission of $L_\alpha\gtrsim10^{41}$erg~s$^{-1}$.

\paragraph*{Observed Ly$\alpha$ equivalent width EW$_\alpha$:}

We compute the Ly$\alpha$ equivalent width EW$_\alpha$ from $L_\alpha$ and the observed UV continuum luminosity at $1500$\AA ($L_c$). The trend of the median EW$_\alpha$ with halo mass follows that of $L_\alpha$, with median EW$_\alpha$ values ranging from $\sim5-30$\AA~ for galaxies in $M_h\simeq10^{10}\msun$ halos to $\sim25-100$\AA~ for galaxies in $M_h\simeq10^{11.3}\msun$ halos. 
More massive galaxies with a strongly attenuated UV continuum -- the fraction of these galaxies increases towards higher halo masses due to the higher abundance of dust -- and high $L_\alpha$ values show EW$_\alpha$ values up to $\sim300$\AA~ in the {\it Clumpy} and {\it Porous} models. However, these high EW$_\alpha$ values are not present in the {\it Gaussian} model for the following reason: in this model, the escape of Ly$\alpha$ and UV continuum radiation differs just by a constant factor, while the dust attenuation of Ly$\alpha$ and UV continuum photons within a galaxy are not only linked via the dust mass in the {\it Clumpy} and {\it Porous} models.

\paragraph*{}

In summary, we find that only more massive galaxies ($M_h\gtrsim10^{10}\msun$) where star formation is not substantially suppressed by SN and radiative feedback from reionisation show significant Ly$\alpha$ emission of $L_\alpha\gtrsim10^{41}$erg~s$^{-1}$. This limitation of observable Ly$\alpha$ emission to more massive galaxies allows the $f_\mathrm{esc}$-dependency of the intrinsic Ly$\alpha$ luminosity to be compensated by a weaker or stronger attenuation of Ly$\alpha$ by dust within a galaxy. If less massive galaxies were visible in Ly$\alpha$, they would break this degeneracy as they would not contain enough dust to attenuate the Ly$\alpha$ radiation in all scenarios sufficiently.

\section{The spatial distribution of Ly$\alpha$ emitting galaxies}
\label{sec_spatial_distribution_LAEs}

\begin{figure*}
    \centering
    \includegraphics[width=0.99\textwidth]{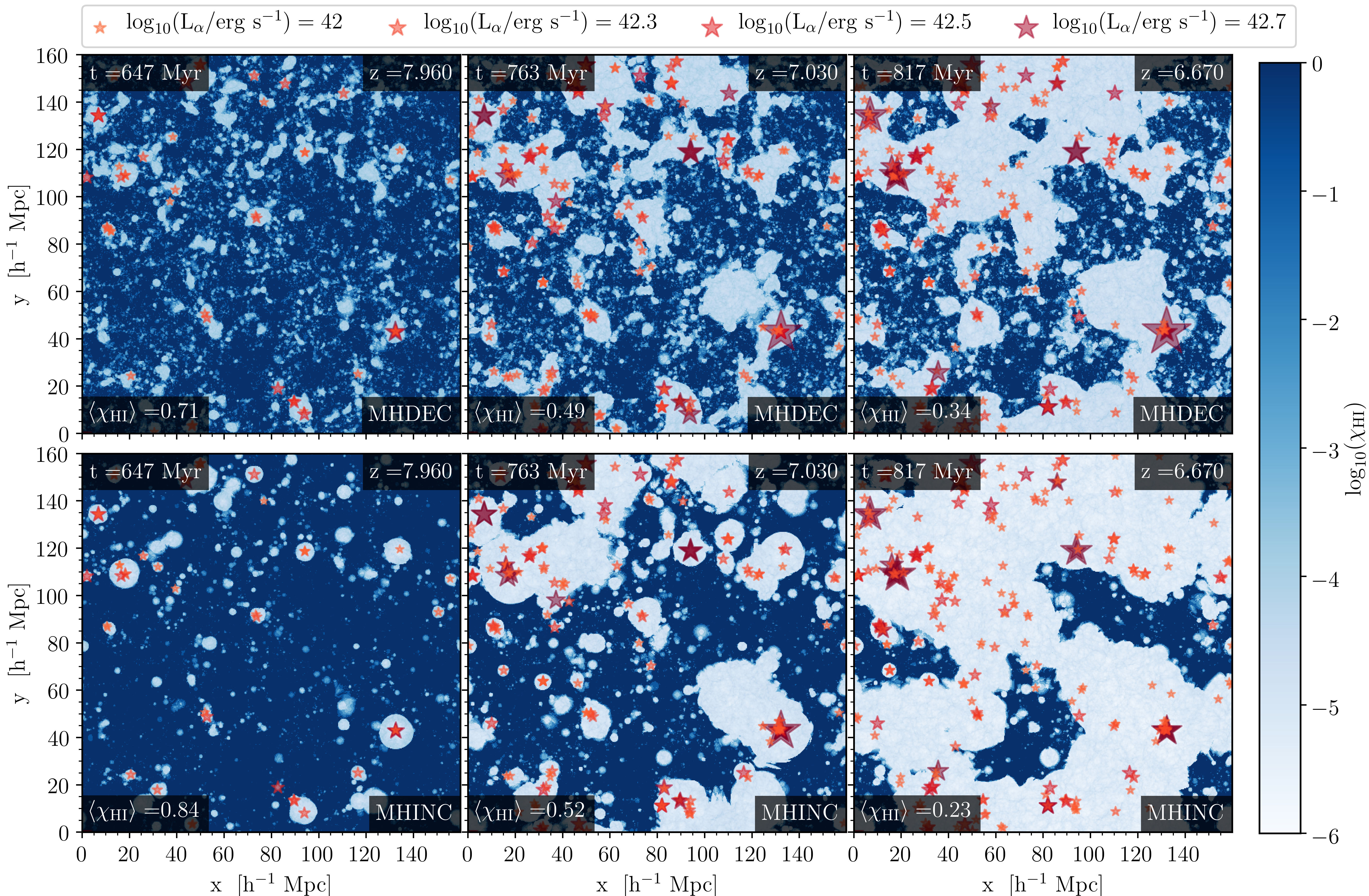}
    \caption{Neutral hydrogen fraction fields at $z=8.0$ (left), $z=7.0$ (centre), and $z=6.6$ (right) for the {\sc mhdec} (top) and {\sc mhinc} {\it Porous} models (bottom). We show a $1.6h^{-1}$cMpc-thick (5 cells) slice through the centre of the simulation box. The blue color scale depicts the volume-averaged value of the neutral fraction in each cell. Red stars show the location of LAEs, with their sizes and colour scale encoding the observed Ly$\alpha$ luminosity along the $z$-direction.}
    \label{fig_XHImaps_with_LAEs}
\end{figure*}

\begin{figure*}
    \centering
    \includegraphics[width=0.99\textwidth]{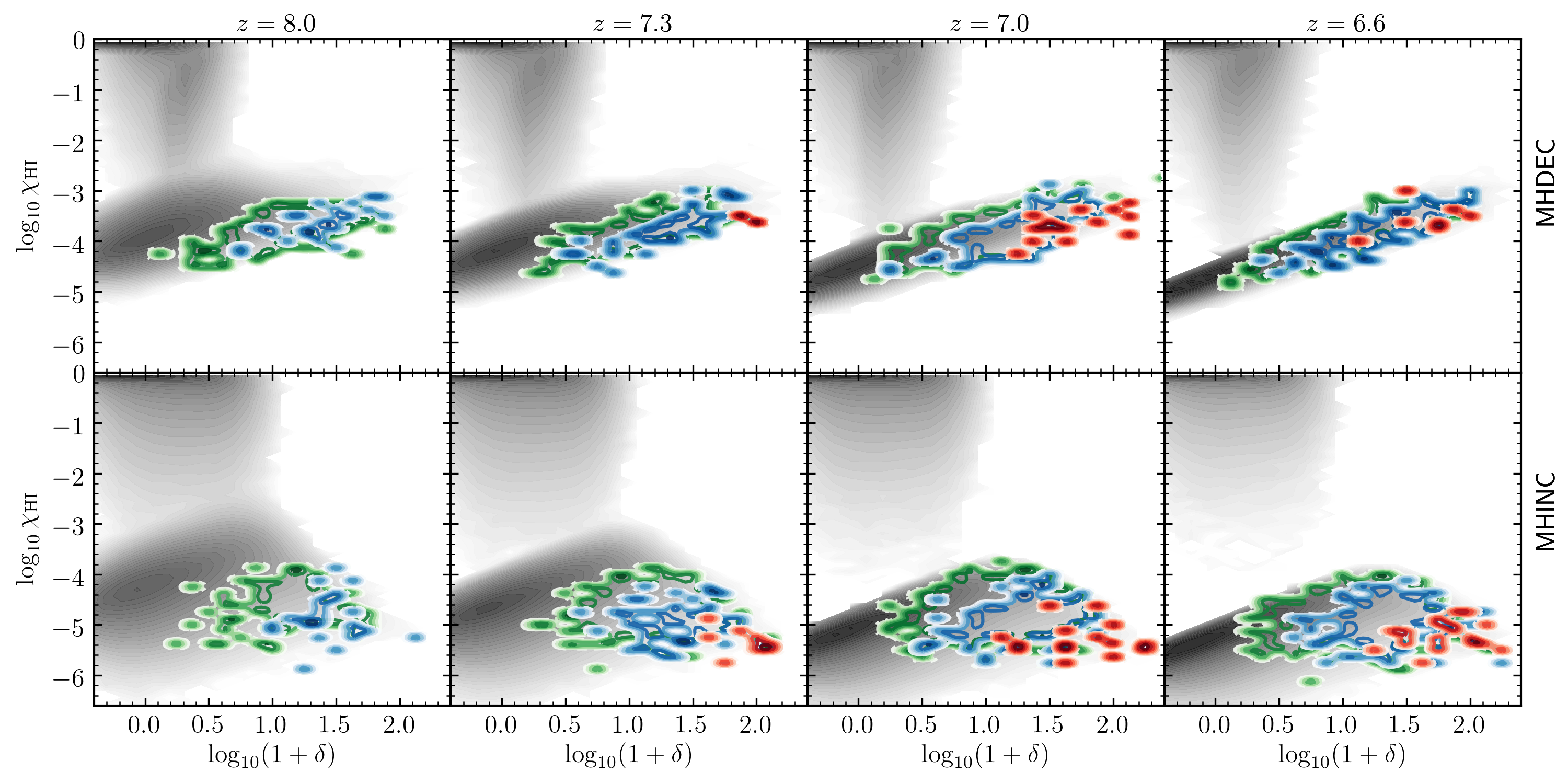}
    \caption{2D probability distribution in $\chi_\mathrm{HI}$ and overdensity for all simulation cells (grey) and galaxies with $L_\alpha\geq10^{42}$erg~s$^{-1}$ (green), $L_\alpha\geq10^{42.5}$erg~s$^{-1}$ (blue), and $L_\alpha\geq10^{43}$erg~s$^{-1}$ (red) in the {\it Porous} model. The top (bottom) row shows results for the reionisation scenario where $f_\mathrm{esc}$ decreases (increases) with halo mass $M_h$.}
    \label{fig_Lya_obs_XHI-DENS}
\end{figure*}

\begin{figure*}
    \centering
    \includegraphics[width=0.99\textwidth]{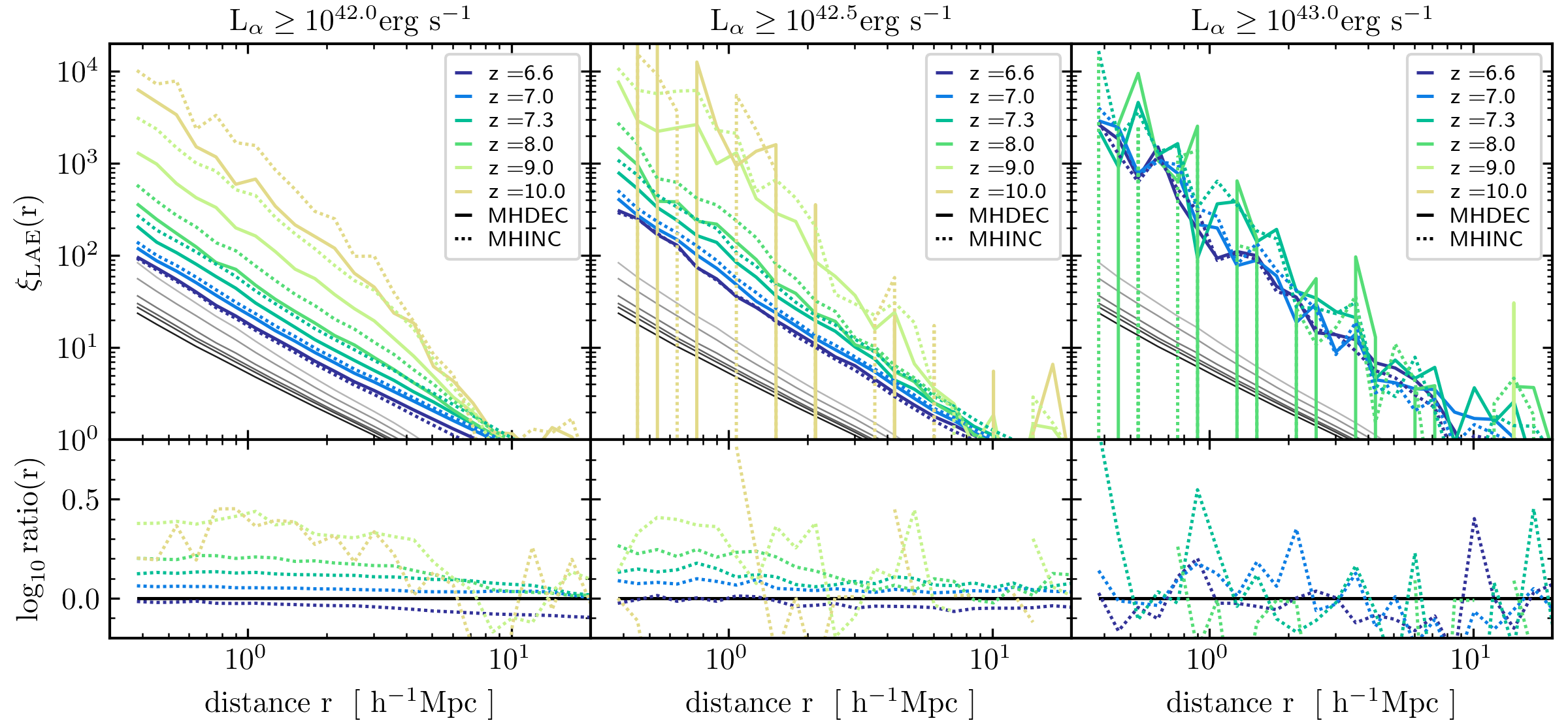}
    \caption{Top panels: 3D correlation function of galaxies that exceed an observed Ly$\alpha$ luminosity of $L_\alpha>10^{42}$erg~s$^{-1}$ (left), $L_\alpha>10^{42.5}$erg~s$^{-1}$ (centre) and $L_\alpha>10^{43}$erg~s$^{-1}$ (right) at $z=10$, $9$, $8$, $7.3$, $7$, $6.6$ for the {\it Porous} model. Solid (dashed dotted) lines show results for the reionisation scenario where $f_\mathrm{esc}$ decreases (increases) with halo mass $M_h$ and assumes $\tau_\mathrm{0,cl}=4\times10^5$ ($2\times10^5$). The grey to black lines indicate the corresponding LBG ($M_\mathrm{UV}<-17$) 3D correlation functions from $z=10$ to $6.6$. Bottom panels: Ratio between the 3D LAE correlation functions of the {\sc mhinc} and the {\sc mhdec} scenario at fixed redshifts.}
    \label{fig_3Dcorrfuncs_FESC}
\end{figure*}

In this Section, we analyse where galaxies with observable Ly$\alpha$ emission are located in the large-scale structure and how their environment and Ly$\alpha$ luminosity distributions differ in our reionisation scenarios ({\sc mhdec} and {\sc mhinc}). For this purpose, we discuss the environment of Ly$\alpha$ emitting galaxies in terms of their large-scale spatial distribution (Fig. \ref{fig_XHImaps_with_LAEs}), their surrounding over-density ($1+\delta$) and \HI fraction ($\chi_\mathrm{HI}$) (Fig. \ref{fig_Lya_obs_XHI-DENS}), and their 3D autocorrelation functions (Fig. \ref{fig_3Dcorrfuncs_FESC}). 
As we yield very similar results for our three Ly$\alpha$ lines profile, we use the {\it Porous} model as a representative case.

\subsection{The environment}

Before detailing the location of Ly$\alpha$ emitting galaxies in the large-scale matter distribution, we briefly discuss the ionisation structure of the IGM using Fig. \ref{fig_XHImaps_with_LAEs} and \ref{fig_Lya_obs_XHI-DENS}. Fig. \ref{fig_XHImaps_with_LAEs} shows the ionisation fields at $z=8$, $7$ and $6.7$ for the {\sc mhdec} (top) and {\sc mhinc} scenarios (bottom). As can be seen in this Figure, if $f_\mathrm{esc}$ decreases with halo mass ({\sc mhdec} scenario), reionisation is not only more extended but also ionised regions are on average smaller, follow more the large-scale density distribution and thus have less bubble-like shapes than if $f_\mathrm{esc}$ increases with halo mass ({\sc mhinc} scenario). The grey contours in Fig. \ref{fig_Lya_obs_XHI-DENS}, showing the two-dimensional probability density distribution of the \HI fraction ($\chi_\mathrm{HI}$) and over-density of the IGM at $z=8$, $7.3$, $7$ and $6.7$ (derived from all cells of the $512^3$ ionisation and density grids output by {\sc astraeus}), complement the picture.
These contours indicate that not only an increasing fraction of the volume becomes ionised as reionisation progresses (from right to left) but also the $\chi_\mathrm{HI}$ values in ionised regions decrease (e.g. from $\chi_\mathrm{HI}\simeq10^{-4}$ ($10^{-4.3}$) in average dense regions with $\log_{10}(1+\delta)+1$ at $z=8$ to $\chi_\mathrm{HI}\simeq10^{-4.7}$ ($10^{-5.2}$) at $z=6.7$ for the {\sc mhdec} ({\sc mhinc}) scenario). The latter is because as galaxies grow in mass with decreasing redshift, their emission of ionising photons increases, leading to a rise of the photoionisation rates within ionised regions and thus lower $\chi_\mathrm{HI}$ values. Moreover, at the same time, as the photoionisation rate within ionised regions becomes increasingly homogeneous, the enhanced number of recombinations in denser regions \citep[for the detailed modelling description see][]{Hutter2018a} leads to a positive correlation between the \HI fraction and density in ionised regions. 
However, the exact value of the photoionisation rate within ionised regions and its spatial distribution depends strongly on the ionising emissivities escaping from the galaxies into the IGM. If less clustered low-mass galaxies drive reionisation -- as in the {\sc mhdec} scenario (top row in Fig. \ref{fig_Lya_obs_XHI-DENS}) --, the resulting photoionisation rate is more homogeneous and lower than if the more strongly clustered massive galaxies are the main drivers of reionisation (c.f. {\sc mhinc} scenario in the bottom row of Fig. \ref{fig_Lya_obs_XHI-DENS}). 
The difference in the photoionisation rate's magnitude explains the shift of the $\chi_\mathrm{HI}$ values by an order of magnitude to lower values in under-dense to moderately over-dense regions ($\log_{10}(1+\delta)\lesssim1.2$) when going from the {\sc mhdec} to the {\sc mhinc} scenario. In contrast, the more inhomogeneous distribution of the photoionisation rate's values enhances this drop in $\chi_\mathrm{HI}$ in over-dense regions where the most massive galaxies are located.

As we can see from the red stars in Fig. \ref{fig_XHImaps_with_LAEs} and coloured contours in Fig. \ref{fig_Lya_obs_XHI-DENS}, galaxies emitting Ly$\alpha$ luminosities of $L_\alpha\geq10^{42}$erg/s always lie in ionised regions for our Ly$\alpha$ line profile models. Although these galaxies trace the ionisation topology, their populations (and thus locations) hardly differ for our two opposing reionisation scenarios. This absence of a significant difference is due to their massive nature \citep[see also e.g.][]{Kusakabe2018}: hence, all Ly$\alpha$ emitting galaxies lie in over-dense regions, with the ones brighter in Ly$\alpha$ located in denser regions (c.f. green to blue to red contours). The latter trend is mainly because more massive galaxies, which exhibit higher star formation rates and produce more ionising and Ly$\alpha$ radiation, are located in denser regions.

\subsection{The clustering}

In this Section, we address the question whether the Ly$\alpha$ luminosity-dependent distribution of LAEs could differ for reionisation scenarios with opposing trends of $f_\mathrm{esc}$ with halo mass. For this purpose, we analyse the 3D autocorrelation function for LAE samples with different minimum Ly$\alpha$ luminosities (Fig. \ref{fig_3Dcorrfuncs_FESC}). We define a galaxy to be an LAE if it has an observed Ly$\alpha$ luminosity of $L_\alpha\geq10^{42}$erg~s$^{-1}$.

Before we discuss the differences between our opposing $f_\mathrm{esc}$ descriptions, we give a brief overview of the global trends and their origins. Firstly, as predicted by hierarchical structure formation, all autocorrelation functions in Fig. \ref{fig_3Dcorrfuncs_FESC} decrease from small to large scales, implying stronger clustering of galaxies on small scales than on large scales. Secondly, the dropping amplitude of the LAE autocorrelation functions with decreasing redshift (from ochre to blue lines) reflects the growth and increasing ionisation of ionised regions. Thirdly, since the $L_\alpha$ value of a galaxy is strongly correlated to its halo mass in our galaxy evolution model, selecting galaxies with increasingly brighter Ly$\alpha$ luminosities (left to right in Fig. \ref{fig_3Dcorrfuncs_FESC}) corresponds to selecting more massive galaxies. The latter explains the increasing amplitude and stronger clustering. Comparing the correlation functions of the $L_\alpha$ selected galaxies with those of LBGs (galaxies with $M_\mathrm{UV}\geq-17$) shows that the Ly$\alpha$ selected galaxies are more massive than our LBGs (solid grey lines). It also shows that the decrease in the clustering of LAEs is partially due to galaxies of a given mass becoming a less biased tracer of the underlying density field as the density of the Universe drops with decreasing redshift.

Comparing the autocorrelation functions of our two opposing $f_\mathrm{esc}$ descriptions, we find that the {\sc mhinc} scenario (dotted lines) has higher autocorrelation amplitudes than the {\sc mhdec} scenario (solid lines) throughout reionisation and for all minimum Ly$\alpha$ luminosities studied. This difference decreases towards larger scales. The reason for these higher amplitudes is twofold: 
On the one hand, the {\sc mhinc} scenario has a lower global average ionisation fraction at $z\gtrsim7$ than the {\sc mhdec} scenario (see Fig. \ref{fig_hist_ion}). Its ionised regions are located around more biased tracers of matter, i.e. more massive galaxies, leading to a stronger clustering. While the scenarios' difference in $\langle\chi_\mathrm{HI}\rangle$ reaches its maximum with $\sim0.13$ around $z\simeq8$, the difference in the autocorrelation amplitudes rises even towards higher redshifts. This is because, with increasing redshift, galaxies of the same mass become more biased tracers of the underlying matter distribution. Thus, the same difference in $\langle \chi_\mathrm{HI}\rangle$ at higher $\langle \chi_\mathrm{HI}\rangle$ values leads to a larger difference in the clustering of LAEs, since the Ly$\alpha$ luminosity of a galaxy correlates strongly with its halo mass. We note that selecting LAEs with a higher minimum Ly$\alpha$ luminosity also corresponds to selecting more biased tracers and yields higher correlation amplitudes (c.f. different panels in Fig. \ref{fig_3Dcorrfuncs_FESC}). 
On the other hand, during the early stages of reionisation, ionised regions grow preferentially around the most biased tracers of the underlying matter field (most massive galaxies) in the {\sc mhinc} scenario. Thus, we would expect that, at the same $\langle\chi_\mathrm{HI}\rangle$ value, LAEs in this scenario are more clustered than LAEs in the {\sc mhdec} scenario where the $f_\mathrm{esc}$ decreasing with rising halo mass counteracts the biased growth of ionised regions. 
Indeed, at $z\lesssim7$, the correlation amplitude in the {\sc mhinc} scenario is higher or similar than in the {\sc mhdec} scenario, although the Universe is similarly or more ionised in the former, respectively. This difference becomes more apparent as we consider higher minimum Ly$\alpha$ luminosities of $L_\alpha>10^{42.5}$erg~s$^{-1}$. It is driven by the higher photoionisation rates in the ionised regions around massive galaxies.

We conclude that, since LAEs coincide with the most massive galaxies located in dense and ionised regions, their clustering is primarily a tracer of the global ionisation state of the IGM. While the exact ionisation topology at fixed $\langle\chi_\mathrm{HI}\rangle$ values has only a secondary effect on the clustering of LAEs during the second half of reionisation, the spatial distribution of LAEs provides a relatively robust tool to map the detailed ionisation history at early times.

\section{The relation of LAEs to LBGs}
\label{sec_LAE_LBG_relation}

\begin{figure*}
    \centering
    \includegraphics[width=0.99\textwidth]{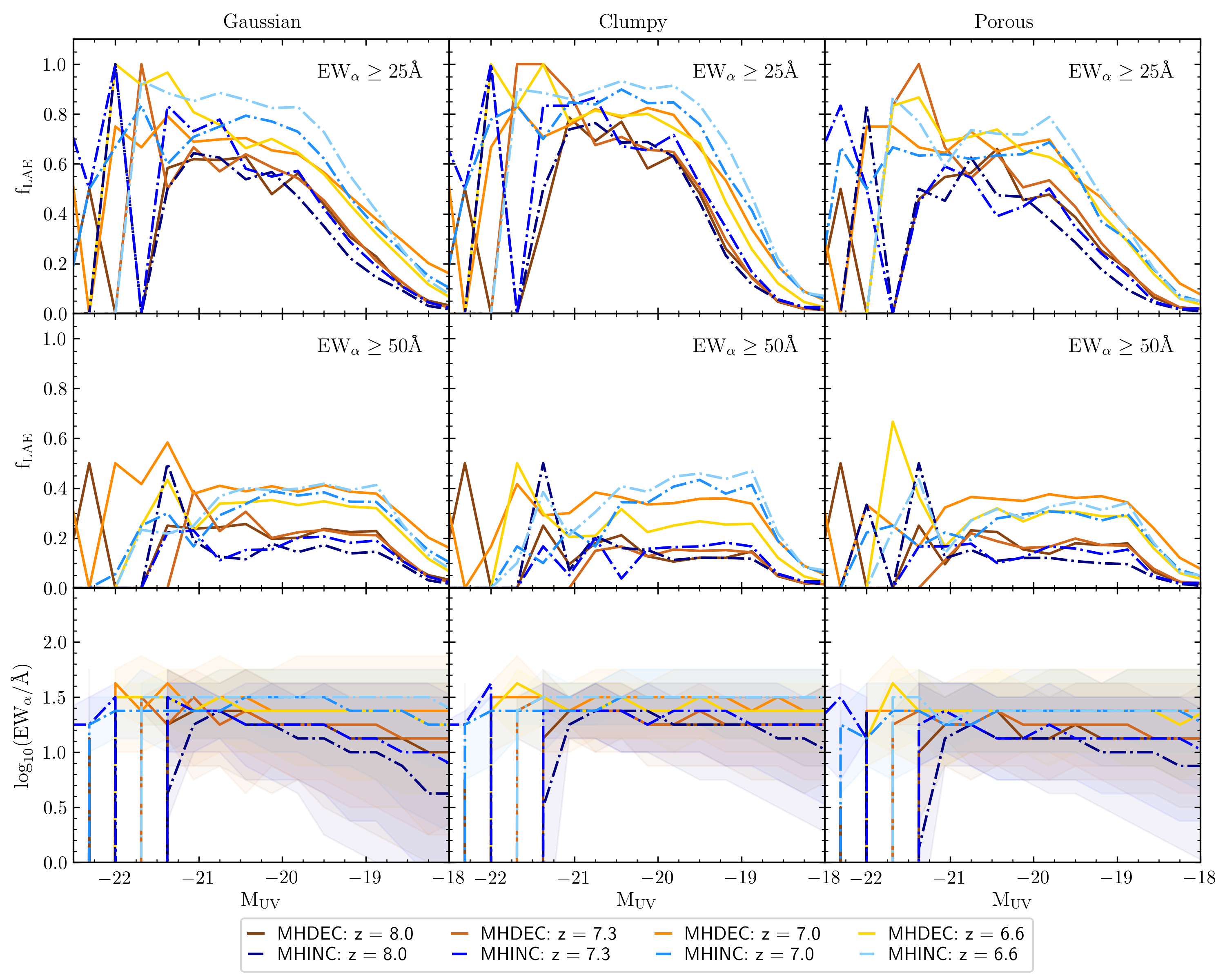}
    \caption{The two top rows depict the fraction of LBGs showing Ly$\alpha$ emissions with $L_\alpha\geq10^{42}$erg~s$^{-1}$ and $\mathrm{EW_\alpha}$ exceeding the value marked in the panels for the different Ly$\alpha$ line profile models as marked. The bottom row shows the corresponding medians of the $\mathrm{EW_\alpha}$ values (lines) and their $\sim1.3\sigma$ uncertainties (shaded regions). Solid (dashed dotted) lines show results for the reionisation scenario where $f_\mathrm{esc}$ decreases (increases) with halo mass $M_h$.}
    \label{fig_LAE_LBG}
\end{figure*}

In this Section, we address the question of what defines whether an LBG shows Ly$\alpha$ emission and why the fraction of LBGs with observable Ly$\alpha$ emission changes as the observed UV continuum luminosity (at $1500$\AA) or the minimum Ly$\alpha$ equivalent width, EW$_\alpha$, rise. For this purpose, we show both the fraction of LBGs with a Ly$\alpha$ equivalent width of at least EW$_\alpha\geq25$\AA~ (top row) and EW$_\alpha\geq50$\AA~ (central row) and the median EW$_\alpha$ value (bottom row) as a function of the UV continuum luminosity in Fig. \ref{fig_LAE_LBG}. 

For our three different Ly$\alpha$ line profile models, we find the median EW$_\alpha$ to exhibit similar values of $\sim4-40$\AA~ at all redshifts shown. Furthermore, the EW$_\alpha$ values range to lower values as galaxies become UV fainter. As galaxies become less massive, this spread in EW$_\alpha$ values reflects the increasingly broader range of star formation rate values to lower values, which traces back to the larger variety of mass assembly histories that increasingly include progenitors with particularly SN feedback-suppressed star formation. The $M_\mathrm{UV}$-dependency of the fraction of LBGs with Ly$\alpha$ emission ($f_\mathrm{LAE}$) also reflects this shift towards lower EW$_\alpha$ values (c.f. top and central row of Fig. \ref{fig_LAE_LBG}): firstly, $f_\mathrm{LAE}$ decreases towards lower UV luminosities, and secondly, this decrease is stronger for lower than higher EW$_\alpha$ cuts. These trends imply that UV bright galaxies are more likely to show higher EW$_\alpha$ values for all our Ly$\alpha$ line profile models and reionisation scenarios. For example, while only $<20\%$ of galaxies with $M_\mathrm{UV}\simeq-18$ exceed EW$_\alpha>25$\AA, $>40\%$ of galaxies with $M_\mathrm{UV}\gtrsim-20$ exceed EW$_\alpha>25$\AA~ and $>5\%$ even EW$_\alpha>50$\AA.

Moreover, both the slight rise of EW$_\alpha$ and $f_\mathrm{LAE}$ values with decreasing redshift and their variation among our different reionisation scenarios can be attributed to the increasing fraction of Ly$\alpha$ radiation that is transmitted through the IGM as the Universe becomes more ionised (see $T_\alpha$ in Fig. \ref{fig_histograms}). For example, at a given redshift $z>7$ ($z<7$), the EW$_\alpha$ and $f_\mathrm{LAE}$ values are on average higher (lower) in the {\sc mhdec} than in the {\sc mhinc} scenario, which is due to a more (less) ionised IGM. Similarly, the lower EW$_\alpha$ values reached in the {\it Gaussian} model for UV fainter galaxies are due to the stronger absorption of Ly$\alpha$ radiation by \HI in the IGM. Finally, we note that since in the {\it Clumpy} and {\it Porous} models the attenuation of the UV continuum and Ly$\alpha$ by dust do not necessarily correlate with each other (as e.g. parts of Ly$\alpha$ can escape via random walk), a few galaxies that are attenuated strongly in the UV but less in Ly$\alpha$ show high EW$_\alpha$ values of $\sim1000$\AA. Thus, the main driver of these high EW$_\alpha$ values is the dust attenuation of the UV continuum assumed in our models.

Comparing our fraction of LBGs showing Ly$\alpha$ emission with those obtained in observations \citep[e.g.][]{Schenker2012, Schenker2014, Caruana2014, Pentericci2014, Pentericci2018, Mason2019}, we find that (i) the observed trend of $f_\mathrm{LAE}$ decreasing towards higher UV luminosity agrees roughly with our results for EW$_\alpha>50$\AA~ but not for EW$_\alpha>25$\AA, and (ii) our $f_\mathrm{LAE}$ values are higher than those inferred from observations (again more so for EW$_\alpha>25$\AA~ than EW$_\alpha>50$\AA). These discrepancies hint either at our model predicting too high Ly$\alpha$ or too low UV luminosities (particularly for more massive galaxies) despite reproducing the observed Ly$\alpha$ and UV LFs, or observations missing bright LAEs. Interestingly, we find that the fraction of LBGs with high EW$_\alpha$ values of $f_\mathrm{LAE}(\mathrm{EW}_\alpha>100$\AA$)\simeq1-12\%$ and $f_\mathrm{LAE}(\mathrm{EW}_\alpha>240$\AA$)\lesssim1\%$ in the {\it Clumpy} and {\it Porous} models are in rough agreement with the results from deep MUSE observations at $z=3-6$ that consider only LAEs with detected UV continuum \citep{Kerutt2022}. A higher abundance of high EW$_\alpha$ values has been found in various high-redshift LAE observations \citep[e.g.][]{Shibuya2018, Malhotra_Rhoads2002, shimasaku2006}.
Nevertheless, our $f_\mathrm{LAE}$ values agree roughly with the results from radiative hydrodynamical simulations post-processed with Ly$\alpha$ radiative transfer, such as {\sc sphinx} \citep[c.f. Fig. B1 in][]{Garel2021}. 

\section{Conclusions}
\label{sec_conclusions}

We apply our new framework for LAEs to different reionisation scenarios, and analyse how the escape fraction of \HI ionising photons, $f_\mathrm{esc}$, and its dependence on halo mass affect the luminosity-dependent number and spatial distributions of LAEs. Besides $f_\mathrm{esc}$ affecting the IGM ionisation topology and the strength of the Ly$\alpha$ line produced in the ISM, its sensitivity to the density and velocity structure of ISM gas and dust has been found to correlate with the Ly$\alpha$ line profile emerging from a galaxy and the fraction of Ly$\alpha$ radiation escaping into the IGM. Notably, the emerging Ly$\alpha$ line profile reflects the attenuation by dust in the ISM and can also change the fraction of Ly$\alpha$ radiation that traverses the IGM unattenuated by \HI. For this reason, we build an analytical model for Ly$\alpha$ line profiles that emerge from a Ly$\alpha$ source surrounded by a shell of dusty gas clumps interspersed with low-density channels. Our model reproduces the numerical radiative transfer results of a shell with dust gas clumps of different sizes as presented in \citet{Gronke2017}.
By coupling this model to {\sc astraeus}, a semi-numerical model coupling galaxy evolution and reionisation self-consistently, we derive the Ly$\alpha$ line profiles emerging from the simulated galaxy population and explore the resulting large-scale distribution of LAEs for different dependencies of $f_\mathrm{esc}$ on halo mass (decreasing, constant, increasing) and Ly$\alpha$ line profiles (Gaussian profile, shell of dusty clumps interspersed with low-density channels or not). For this parameter space, we analyse the resultant ionisation topologies, the dependencies of Ly$\alpha$ line profiles and Ly$\alpha$ properties on halo mass, and the location of galaxies with observable Ly$\alpha$ emission in the large-scale structure. 
Our main results are the following:

\begin{enumerate}
    \item For a shell consisting of clumps of the same size, the Ly$\alpha$ line profile emerging from a galaxy develops from a central peak at the Ly$\alpha$ resonance dominated to a double peak dominated profile as it becomes more massive. Adding low-density channels results in either a weakening of this trend, particularly as $f_\mathrm{esc}$ increases with rising halo mass.
    \item In all reionisation scenarios and Ly$\alpha$ line profile models, LAEs (galaxies with $L_\alpha\geq10^{42}$erg~s$^{-1}$) are more massive galaxies with $M_h\gtrsim10^{10}\msun$. These galaxies exhibit continuous star formation and are biased tracers of the underlying mass density distribution. Both allow efficient transmission of the Ly$\alpha$ line through the IGM by facilitating the build-up of ionised regions around them. In contrast, less massive galaxies are surrounded by smaller ionised regions, which results in their Ly$\alpha$ radiation being significantly attenuated by \HI in the IGM.
    \item As LAEs are more massive galaxies and the most biased tracers of the underlying mass density distribution, they are located in the densest and most highly ionised regions. This finding holds for any inside-out reionisation scenario where dense regions containing massive galaxies are ionised before under-dense voids and for Ly$\alpha$ line profiles exhibiting emission around/close to the Ly$\alpha$ resonance \citep[see also][]{Hutter2014, Hutter2017}. In such scenarios, the spatial distribution of LAEs is primarily sensitive to the global ionisation fraction and only in second-order to the ionisation topology or the trend of $f_\mathrm{esc}$ with halo mass.
    \item As the observable Ly$\alpha$ LFs are composed of the Ly$\alpha$ emission from more massive galaxies, a decrease in their intrinsic Ly$\alpha$ luminosities (Ly$\alpha$ produced in the ISM) due to higher $f_\mathrm{esc}$ values can be compensated by reducing the attenuation by dust in the ISM \citep[echoing the degeneracy found in][]{Hutter2014}. However, if $f_\mathrm{esc}$ exceeds $\sim0.5$ for the most massive galaxies ($M_h\gtrsim10^{11}\msun$), their intrinsic Ly$\alpha$ luminosity is too low to reproduce the observed Ly$\alpha$ LFs \citep[see also][]{Hutter2014}.
\end{enumerate}

All combinations of our reionisation scenarios and Ly$\alpha$ line profile models result in Ly$\alpha$ and UV luminosities in reasonable agreement with observational constraints. However, although two of the three Ly$\alpha$ line profile models investigated use parameterisations of numerical Ly$\alpha$ radiative transfer simulation results, they represent idealised scenarios where the gas in each galaxy is distributed in clumps of the same mass. In reality, the density and velocity distributions of gas and dust in the ISM are more complex:
Firstly, the dusty gas clumps will have different masses, with a distribution close to that of a scale-free one at the massive end. Such a mass distribution would result in more massive galaxies having larger clumps than less massive galaxies, which again would lead to a homogenisation of their Ly$\alpha$ line profiles where more massive (less massive) galaxies have an enhanced (weakened) central peak component and a weakened (enhanced) double-peak component. This change in the Ly$\alpha$ line profiles would result in the Ly$\alpha$ radiation being less (more) attenuated by dust in the ISM and traversing the IGM more (less) efficiently. 
Secondly, the medium between the clumps as well as the low-density channels might not be fully ionised (and very unlikely to be gas-free), causing the Ly$\alpha$ radiation escaping close to its resonance (central peak used in this work) to contribute to a narrower double-peak profile. Additionally, the gas may exhibit a turbulent velocity structure that could broaden the double-peak component. Both partially neutral low-density channels and an inhomogeneous velocity structure are likely to enhance the transmission of Ly$\alpha$ through the IGM.
Thirdly, the attenuation of Ly$\alpha$ radiation by dust in the ISM depends on the distribution of dust in clumps. While our model assumes that gas and dust are perfectly mixed, a scenario where dust condensates in the centre surrounded by a shell of hydrogen gas would lower the absorption probability per clump and enhance the escape fraction of Ly$\alpha$ photons from a galaxy.
Finally, simulations and observations of local analogues of high-redshift galaxies (i.e. regarding their extreme metallicity and ionisation continuum properties) indicate that stellar feedback, especially that of supernovae, heat the gas and drive gas outflows \citep[e.g.][]{Gronke2020, Kakiichi2021, Carr2021, Fielding2022, Xu2023, Hu2023}. Indeed expanding homogeneous shell models have been used to fit observed Ly$\alpha$ profiles from high-redshift analogues \citep[e.g.][]{Gronke2017, Orlitova2018}, however, the inferred outflow velocities are on average lower than those inferred from ultraviolet absorption lines of low-ionisation-state elements \citep{Orlitova2018, Xu2023}, hinting at more complex outflow geometries and kinematics of the neutral gas \citep[see e.g.][]{Carr2021, Blaizot2023}. In general, outflowing neutral gas causes the Ly$\alpha$ photons to redshift, enabling easier escape from the galaxy and transmission through the IGM. While in principle outflows could enhance the observed Ly$\alpha$ emission, particularly from low-mass ($M_h\lesssim10^{10}\msun$) galaxies, and make the large-scale LAE distribution more sensitive to the ionisation topology, their velocities or neutral gas fraction might be not sufficient to redshift the Ly$\alpha$ radiation out of absorption. In future work, we will extend our analytical Ly$\alpha$ line models towards more realistic outflow geometries and kinematics and explore whether this will affect the large-scale LAE distribution during reionisation significantly.

Our Ly$\alpha$ line profile models, despite being limited by the simplified structure assumed for the ISM, represent a first step towards more complex analytical models for the Ly$\alpha$ line emerging from galaxies that are computationally efficient enough to derive the Ly$\alpha$ emitter populations in large cosmological simulations. To date, many models deriving the large-scale distribution of LAEs assume Ly$\alpha$ line profiles that arise from outflowing gas, consisting of a dominant red and a negligible blue peak \citep{Mesinger2015, Mason2018, Weinberger2019}. However, such profiles are hardly seen in Ly$\alpha$ radiative transfer simulations of simulated galaxies \citep[see e.g.][]{Laursen2011, Garel2021, Blaizot2023}.

Finally, our finding that the spatial distribution of LAEs is not sensitive to the dependence of $f_\mathrm{esc}$ with halo mass suggests that LAEs alone can not help to constrain any gradual dependence of $f_\mathrm{esc}$ with galactic properties. Any dependency introduced in the intrinsic Ly$\alpha$ luminosity can be compensated by the opposed trend of the Ly$\alpha$ escape fraction, achieved by changing the ISM gas and dust distribution. This insensitivity to $f_\mathrm{esc}$  dependencies makes LAEs relatively robust tracers of the underlying density field that we can use to pin down the ionisation topology. Constraining $f_\mathrm{esc}$ during the EoR will require a combination of ionisation topology measurements through the \HI 21cm signal and measurements of other emission lines. 

\section*{Acknowledgements}

We thank Max Gronke and Peter Laursen for useful discussions and the anonymous referee for their comments. AH, GY, LL, PD and SG acknowledge support from the European Research Council's starting grant ERC StG-717001 (``DELPHI"). AH, MT, PD also acknowledge support from the NWO grant 016.VIDI.189.162 (``ODIN") and the European Commission's and University of Groningen's CO-FUND Rosalind Franklin program. AH acknowledges support by the VILLUM FONDEN under grant 37459. PD thanks the Institute for Advanced Study (IAS) Princeton, where a part of this work was carried out, for their generous hospitality and support through the Bershadsky Fund. 
GY acknowledges financial support from  MICIU/FEDER under project grant PGC2018-094975-C21. 
We thank Peter Behroozi for creating and providing the {\sc rockstar} merger trees of the {\sc vsmdpl} and {\sc esmdpl} simulations. The authors wish to thank V. Springel for allowing us to use the L-Gadget2 code to run the different Multidark simulation boxes, including the {\sc vsmdpl} and {\sc esmdpl} used in this work. The {\sc vsmdpl} and {\sc esmdpl} simulations have been performed at LRZ Munich within the project pr87yi. The authors gratefully acknowledge the Gauss Centre for Supercomputing e.V. (www.gauss-centre.eu) for funding this project by providing computing time on the GCS Supercomputer SUPERMUC-NG at Leibniz Supercomputing Centre (www.lrz.de). 
The CosmoSim database (\url{www.cosmosim.org}) provides access to the simulations and the Rockstar data. The database is a service by the Leibniz Institute for Astrophysics Potsdam (AIP). 
The Cosmic Dawn Center (DAWN) is funded by the Danish National Research Foundation under grant No. 140.
This research made use of \texttt{matplotlib}, a Python library for publication quality graphics \citep{hunter2007}; and the Python library \texttt{numpy} \citep{numpy}.


\section*{Data Availability}

The source code of the semi-numerical galaxy evolution and reionisation model within the {\sc astraeus} framework is available on GitHub (\url{https://github.com/annehutter/astraeus}). The underlying N-body DM simulation, the {\sc astraeus} simulations and derived data in this research will be shared on reasonable request to the corresponding author.


\bibliographystyle{mnras}
\bibliography{mypapers, general, intro, lya, fesc} 



\appendix

\section{Geometrical correction factor $\xi$}
\label{app_correction_factor}

When deriving the attenuation of the UV continuum by dust, we have assumed the light sources and dust to be homogeneously distributed within a slab. However, the numerical Ly$\alpha$ radiative transfer simulations in \citet{Gronke2017} assume a screen of dust and gas between the light sources and the observer. To make the escape fractions of the UV continuum and Ly$\alpha$ radiation consistent, we introduce a geometrical correction factor that adjusts the Ly$\alpha$ escape fraction for sources behind a screen to that for sources distributed in a dusty gas slab. According to \citet{Forero-Romero2011}, the Ly$\alpha$ escape fraction relations for these two geometries are given by
\begin{eqnarray}
    f_\mathrm{esc}^\mathrm{screen}(\tau_0) &=& \frac{1}{\cosh\left( \xi_0 \sqrt{(a\tau_0)^{1/3} \tau_a} \right)} \\
    f_\mathrm{esc}^\mathrm{slab}(\tau_0) &=& \frac{1 - \exp(-P)}{P} \\
    P &=& \epsilon_0 \left( (a\tau_0)^{1/3} \tau_a \right)^{3/4} \\
    \tau_a &=& \tau_\mathrm{d,total} (1-A),
\end{eqnarray}
where $A$ is the albedo. By equating the expressions for the Ly$\alpha$ escape fractions,
\begin{eqnarray}
f_\mathrm{esc}^\mathrm{screen}(\tau_0^\mathrm{eff}) &=& f_\mathrm{esc}^\mathrm{slab}(\tau_0),
\end{eqnarray}
we derive a correction factor $\xi$ that reduces the \HI column density in the screen geometry to the slab geometry.
\begin{eqnarray}
    \xi &=& \frac{\tau_0^\mathrm{eff}}{\tau_0} = \min\left(\xi_\mathrm{max}, \frac{\epsilon_0}{\xi_0^{3/2}} \frac{\left( \mathrm{arcosh}\left( \frac{P}{1-e^{-P}} \right) \right)^{3/2}}{P}\right) \\
    P &=& \epsilon_0\ a^{1/4} \left( (1-A) \frac{M_\mathrm{d}}{M_\mathrm{HI}} \frac{\kappa_\mathrm{abs} m_\mathrm{H}}{\sigma_\mathrm{HI}} \right)^{3/4} \tau_0
\end{eqnarray}
Here we assume $A=0.5$ and $\xi_0=2.48$, with $\xi_0$ being adjusted to reproduce $f_\mathrm{esc}^\mathrm{screen}$ shown in Fig.~1 in \citet{Forero-Romero2011}. $\xi_\mathrm{max}=0.35$ represents an upper limit for a dust free homogeneous distribution of gas and sources. We derived its value as follows: Firstly we sum the Neufeld solutions (Eqn. \ref{eq_Jslab}) for an equidistant set of $\tau_0$ values between $[0,\tau_0]$. Secondly, from the resulting Ly$\alpha$ line profile we estimate the effective $\tau_0^\mathrm{eff}$ value by measuring the peak positions ($x_p^\mathrm{eff}$). Relating these peak positions to those obtained for the single Neufeld solution for $\tau_0$ ($x_p$), we obtain the ratio $x_p^\mathrm{eff}/x = \left( \tau_0^\mathrm{eff}/\tau_0 \right)^{1/3} = \left( N_\mathrm{HI}^\mathrm{eff}/N_\mathrm{HI} \right)^{1/3}$, and the correction factor $\xi_\mathrm{max}=(x_p^\mathrm{eff}/x_p)^{1/3}$.
We have also checked that applying the correction factor $\xi$ to $N_\mathrm{HI}$ reproduces the correct shift in the Ly$\alpha$ peak positions $x_p$ shown in Fig.~A5 in \citet{Forero-Romero2011}.

\section{Delayed non-bursty supernova feedback scheme}
\label{app_delayed_non-bursty_SNscheme}

We briefly describe our new formalism for the number of SN exploding if the stellar mass formed in a time step is assumed to form at a continuous rate across that time step.
For a given star formation history $\mathrm{SFR}(t)$, the differential number of SN after a time $t$ is given by
\begin{eqnarray}
\frac{\mathrm{d}N_\mathrm{SN}}{\mathrm{d}t}(t) &=& \int_0^{\infty} \mathrm{d}t'\ \mathrm{SFR}(t')\ \nu(t-t').
\label{eq_SNnumber}
\end{eqnarray}
$\nu(t)$ is the differential number of SN per stellar mass formed at $t'=0$ and exploding at $t'=t$, and hence yields as
\begin{eqnarray}
\nu(t) &=& M_\mathrm{SN}^{-\gamma}(t)\ \frac{\mathrm{d}M_\mathrm{SN}}{\mathrm{d}t} \Theta(t-t_\mathrm{\star, high})\ \Theta(t_\mathrm{SN, low}-t),
\label{eq_SNfrac}
\end{eqnarray}
with $\gamma$ being the slope of the assumed IMF, $t_\mathrm{\star, high}$ being the time after which the most massive stars sampled by the IMF, $M_\mathrm{\star,high}$, explode as SN, and $t_\mathrm{SN,low}$ the time that it takes a star with the lowest stellar mass to explode as SN ($M_\mathrm{SN,low}=8\Msun$). Stars of mass $M_\mathrm{SN}$ explode after a time $t$ and the corresponding relation is described by
\begin{eqnarray}
\frac{M_\mathrm{SN}}{\Msun} &=& \left( \frac{t/\mathrm{Myr} -3 }{1.2\times 10^3} \right)^{-1/1.85} = a^{-c} (t - 3)^{-c}. 
\label{eq_SNmassTimeRelation}
\end{eqnarray}
For constant star formation with 
\begin{eqnarray}
\mathrm{SFR}(t)=
\begin{cases}
0 & t<t_i\\
s_0 & t_i\leq t\leq t_f\\
0 & t_f<t
\end{cases}
\end{eqnarray}
we yield after inserting Eqn. \ref{eq_SNfrac} and \ref{eq_SNmassTimeRelation} into Eqn. \ref{eq_SNnumber}
\begin{eqnarray}
\frac{\mathrm{d}N_\mathrm{SN}}{\mathrm{d}t}(t) &=& \int_{t_i}^{t_f} \mathrm{d}t'\ s_0\ \nu(t-t') \\
&=& \int_{t_\mathrm{min}}^{t_\mathrm{max}} \mathrm{d}t'\ s_0 \ M_\mathrm{SN}^{-\gamma}(t-t')\  \frac{\mathrm{d}M_\mathrm{SN}}{\mathrm{d}(t-t')} \nonumber\\
&=& - \int_{t_\mathrm{min}}^{t_\mathrm{max}} \mathrm{d}t'\ s_0 \ c a^{c(\gamma-1)} \left( t-t'-3 \right)^{c(\gamma-1)-1} \nonumber\\
&=& s_0\ \frac{a^{c(\gamma-1)}}{1-\gamma} \nonumber \\
&\times& \left[ (t-t_\mathrm{min}-3)^{c(\gamma-1)} - (t-t_\mathrm{max}-3)^{c(\gamma-1)} \right] \nonumber \\
&=& s_0\ \frac{a^{c(\gamma-1)}}{1-\gamma} \left[ f_\mathrm{min}(t) - f_\mathrm{max}(t) \right] \nonumber
\label{eq_SNnumber_constSF}
\end{eqnarray}
with
\begin{eqnarray}
t_\mathrm{min} &=& \max[t_i, t-t_\mathrm{SN,low}] \\
t_\mathrm{max} &=& \min[t_f, t-t_\mathrm{\star,high}]
\end{eqnarray}
and \begin{eqnarray}
t_\mathrm{max} &\geq& t_\mathrm{min}.
\end{eqnarray}
These relations result in the following additional criteria
\begin{eqnarray}
t &\geq& t_i + t_\mathrm{\star,high} \\
t &\leq& t_f + t_\mathrm{SN,low}.
\end{eqnarray}
Eqn. \ref{eq_SNnumber_constSF} describes the differential number of SN exploding between the onset of star formation ($t_i$) and time $t$ assuming constant star formation from $t_i$ to $t_f$.
However, to obtain the total number of SN exploding in a given time step, i.e. between $t_{j-1}$ and $t_j$, we need to integrate over all contributions from $t_{j-1}\leq t\leq t_j$ (i.e. integrating Eqn. \ref{eq_SNnumber_constSF} over time $t$),
\begin{eqnarray}
N_\mathrm{SN}(t_i, t_f, t_{j-1}, t_j) &=& \int_{t_{j-1}}^{t_j} \mathrm{d}t\  \frac{\mathrm{d}N_\mathrm{SN}}{\mathrm{d}t} \\
&=& \int_{t_{j-1}}^{t_j} \mathrm{d}t\ s_0\ \frac{a^{c(\gamma-1)}}{1-\gamma} \left[ f_\mathrm{min}(t) - f_\mathrm{max}(t) \right]. \nonumber 
\label{eq_totSNnumber_constSF}
\end{eqnarray}
We solve the different summands in the integral separately, yielding
\begin{eqnarray}
F_\mathrm{max} &=& \int_{t_{j-1}}^{t_j} \mathrm{d}t\ f_\mathrm{max}(t) \\
&=& \int_{t_{j-1}}^{t_j} \mathrm{d}t\ (t-t_\mathrm{max}-3)^{c(\gamma-1)} \nonumber \\
&& \times\ \Theta(t_f + t_\mathrm{SN,low} -t)\ \Theta(t-t_i+t_\mathrm{\star,high}) \nonumber \\
&=& \int_{\max(t_{j-1}, t_i+t_\mathrm{\star,high})}^{\min(t_j, t_f+t_\mathrm{\star,high})} \mathrm{d}t\ \left( t_\mathrm{\star,high}-3 \right)^{c(\gamma-1)} \nonumber \\
&&+ \int_{\max(t_{j-1}, t_f+t_\mathrm{\star,high})}^{\min(t_j, t_f+t_\mathrm{SN,low})} \mathrm{d}t\ \left( t-t_f-3 \right)^{c(\gamma-1)} \nonumber \\
&=& \left[ \left( t_\mathrm{\star,high}-3 \right)^{c(\gamma-1)}\ t \right]_{\max(t_{j-1}, t_i+t_\mathrm{\star,high})}^{\min(t_j, t_f+t_\mathrm{\star,high})}  \nonumber \\
&&+ \left[ \frac{\left( t-t_f-3 \right)^{c(\gamma-1)+1}}{c(\gamma-1)+1} \right]_{\max(t_{j-1}, t_f+t_\mathrm{\star,high})}^{\min(t_j, t_f+t_\mathrm{SN,low})} \nonumber
\label{eq_Fmax}
\end{eqnarray}
and 
\begin{eqnarray}
F_\mathrm{min} &=& \int_{t_{j-1}}^{t_j} \mathrm{d}t\ f_\mathrm{min}(t) \\
&=& \int_{t_{j-1}}^{t_j} \mathrm{d}t\ (t-t_\mathrm{min}-3)^{c(\gamma-1)} \nonumber \\
&& \times\ \Theta(t_f + t_\mathrm{SN,low} -t)\ \Theta(t-t_i+t_\mathrm{\star,high}) \nonumber \\
&=& \int_{\max(t_{j-1}, t_i+t_\mathrm{\star,high})}^{\min(t_j, t_i+t_\mathrm{SN,low})} \mathrm{d}t\ \left( t-t_i-3 \right)^{c(\gamma-1)} \nonumber \\
&&+ \int_{\max(t_{j-1}, t_i+t_\mathrm{SN,low})}^{\min(t_j, t_f+t_\mathrm{SN,low})} \mathrm{d}t\ \left( t_\mathrm{SN,low}-3 \right)^{c(\gamma-1)}\nonumber \\
&=& \left[ \left( t_\mathrm{\star,high}-3 \right)^{c(\gamma-1)}\ t \right]_{\max(t_{j-1}, t_i+t_\mathrm{\star,high})}^{\min(t_j, t_i+t_\mathrm{SN,low})}  \nonumber \\
&&+ \left[ \frac{\left( t-t_f-3 \right)^{c(\gamma-1)+1}}{c(\gamma-1)+1} \right]_{\max(t_{j-1}, t_i+t_\mathrm{SN,low})}^{\min(t_j, t_f+t_\mathrm{SN,low})}. \nonumber
\label{eq_Fmin}
\end{eqnarray}
Inserting Eqn. \ref{eq_Fmax} and \ref{eq_Fmin} into Eqn. \ref{eq_totSNnumber_constSF}, we obtain
\begin{eqnarray}
N_\mathrm{SN}(t_i, t_f, t_{j-1}, t_j) &=& s_0\ \frac{a^{c(\gamma-1)}}{1-\gamma} \left[ F_\mathrm{min} - F_\mathrm{max} \right]
\label{eq_totSNnumber_constSF_solved}
\end{eqnarray}

For a given star formation law $\mathrm{SFR}(t)$, the total stellar mass formed across all time is
\begin{eqnarray}
M_\star^\mathrm{tot} &=& \int_{0}^{\infty}\mathrm{d}t\ \mathrm{SFR}(t)\ \left[ \int_{M_\mathrm{\star,high}}^{M_\mathrm{\star,low}} \mathrm{d}m\ m^{1-\gamma} \right] \\
&=& \int_{0}^{\infty}\mathrm{d}t\ \mathrm{SFR}(t)\ \frac{M_\mathrm{\star,low}^{2-\gamma} - M_\mathrm{\star,high}^{2-\gamma}}{2-\gamma}. \nonumber 
\label{eq_totMstar_constSF}
\end{eqnarray}
Hence, for a constant star formation between $t_i$ and $t_f$, we finally obtain
\begin{eqnarray}
M_\star^\mathrm{tot}(t_i,t_f) &=& \frac{s_0}{t_f-t_i} \frac{M_\mathrm{\star,low}^{2-\gamma} - M_\mathrm{\star,high}^{2-\gamma}}{2-\gamma}
\label{eq_totMstar_constSF_solved}
\end{eqnarray}

Finally, from Eqn. \ref{eq_totSNnumber_constSF_solved} and \ref{eq_totMstar_constSF_solved}, we derive the number of SN exploding between times $t_{j-1}$ and $t_j$ from stars formed between $t_i$ and $t_j$ per stellar mass as
\begin{eqnarray}
\frac{N_\mathrm{SN}(t_i, t_f, t_{j-1}, t_j)}{M_\star^\mathrm{tot}(t_i,t_f)} &=& \frac{2-\gamma}{1-\gamma}\ \frac{a^{c(\gamma-1)}}{M_\mathrm{\star,low}^{2-\gamma} - M_\mathrm{\star,high}^{2-\gamma}}\ \frac{F_\mathrm{min}-F_\mathrm{max}}{t_f-t_i} \nonumber \\
\end{eqnarray}


\bsp	
\label{lastpage}
\end{document}